\documentclass[a4paper,fleqn]{cas-sc}
\usepackage[authoryear]{natbib}
\usepackage{longtable}
\usepackage{caption}
\usepackage{graphicx}
\usepackage{float}
\usepackage{subfigure}
\usepackage{threeparttable}
\usepackage[figuresright]{rotating}
\usepackage{amsmath}
\usepackage{amssymb}
\def\tsc#1{\csdef{#1}{\textsc{\lowercase{#1}}\xspace}}
\tsc{WGM}
\tsc{QE}

\begin{document}

\let\WriteBookmarks\relax
\def\floatpagepagefraction{1}
\def\textpagefraction{.001}

\shortauthors{<Peng et al.>}  
\shorttitle{Discovery of four new EL CVn-type binaries}    
\title{Discovery of four new EL CVn-type binaries in the {\it Gaia} eclipsing binaries}

\author[1]{Yuhui Peng}
\affiliation[1]{organization={School of Physics and Astronomy, China West Normal University}, 
            city={Nanchong},
            postcode={637009}, 
            country={China}}

\author[1]{Kun Wang}[orcid=0000-0002-5745-827X]
\cormark[1]
\ead{kwang@cwnu.edu.cn}

\author[1]{Anbing Ren}

\cortext[cor1]{Corresponding author}

\begin{abstract}
In this paper, we performed a search for EL CVn-type binaries based on the {\it Gaia} and {\it TESS} data.
Through the combination of the {\it Gaia} DR3 eclipsing binary catalogue and the {\it Gaia} DR3 spectroscopic binary catalogue, 
we have identified 13 stars exhibiting EL CVn-like characteristics.
Among these stars, nine have already been identified as EL CVn binaries, while the remaining four are newly discovered candidates.
The photometric solutions and absolute parameters of the four binary systems were obtained by utilizing {\it TESS} photometry 
and orbital elements from {\it Gaia} DR3 spectroscopic binary catalogue.
Light-curve modeling has revealed that all four binary systems with a detached configuration exhibit very low mass ratios, approximately $q\simeq$0.1. 
The derived physical parameters indicate that the four binary systems are highly probable candidates for being newly discovered EL CVn-type binaries.
A preliminary frequency analysis was conducted on the residuals of the light curves, after the removal of binarity effects.
The results suggest that the primary component star of TIC 100011519, TIC 219485855 and TIC 464641792 might be a $\delta$ Sct pulsator. 
However, TIC 399725538 does not exhibit any intrinsic pulsations.
\end{abstract}

%%Research highlights
%\begin{highlights}
%\item Performed a search for EL CVn-type binaries based on the {\it Gaia} and {\it TESS} data.
%\item Identified 13 stars exhibiting EL CVn-like characteristics, nine of which have already been identified as EL CVn binaries, while the remaining four are newly discovered candidates.
%\item The photometric solutions and absolute parameters of the four new EL CVn binaries were obtained by utilizing {\it TESS} photometry 
%and orbital elements from {\it Gaia} DR3 spectroscopic binary catalogue.
%\item The primary component star of TIC 100011519, TIC 219485855 and TIC 464641792 might be a $\delta$ Sct pulsator. 
%\end{highlights}


\maketitle

%________________________________________________ sections below
\section{Introduction}\label{intro}
Extremely low-mass white dwarfs (ELM WDs) and their progenitor stars are a class of helium white dwarfs with masses less than $\sim0.3M_\odot$ \citep{Wang2022}. 
ELM WDs are considered as the product of binary star evolution \citep{Istrate2016, Chen2017, Li2019, Chen2021}.
An ELM WD cannot form through the natural evolution of a single star since the age of the universe is not sufficient for a single star to evolve into a helium-core white dwarf.
Up to now, all known ELM WDs were discovered in binary systems that consist of white dwarfs, neutron stars, or A/F-type dwarf stars as their companions (e.g. \citealt{Maxted2014, Mata2020, Wang2020, Brown2022}).
However, a population of single extremely low mass white dwarfs may exist, 
which could be the remaining cores of giant stars whose outer layers were stripped by a companion during a supernova explosion \citep{Justham2009, Wang2010}.

EL CVn-type binaries are a group of post-mass transfer eclipsing binary systems that consist of an A/F-type dwarf star and either an ELM WD or its precursor \citep{Maxted2011, Lee2020AJ}.
EL CVn binaries show a distinctive light curve characterized by a primary eclipse with a "boxy" shape, i.e. steep ingress and egress and a well-defined flat bottom \citep{Maxted2014}.
The secondary eclipse, which occurs due to the transit of the smaller, hotter ELM WD,
has a comparable depth but is shallower than the total eclipse caused by the occultation of the bigger, cooler dwarf star.
Based on the analysis of their light-curve characteristics, nearly a hundred EL CVn binaries have been identified through the WASP, PTF, Kepler, {\it TESS}, and {\it Gaia} surveys.
(e.g. \citealt{Maxted2014, vanRoestel2018, Zhang2017, Wang2020, Gaia2023A&A674A34G}\footnote{Since Gaia Collaboration papers have many authors, we have adjusted the reference format to display them as "Gaia Collaboration".}). 

In this paper, we have used the rich {\it Gaia} photometric data and derived properties, together with the {\it TESS} data, to search for EL CVn systems.
Identifying these systems helps us understand the evolution of binary stars and the mechanisms behind mass transfer. 
The paper is structured as follows:  Section 2 describes the identification of the EL CVn-like stars using the {\it Gaia} and {\it TESS} data. 
Section 3 introduces the photometry and light-curve modeling. 
Section 4 presents the multifrequency analysis.
Finally, the discussion and summary are presented in Section 5.

\section{Target selection}    
With 34 months of astrometry and photometric observations, 
Gaia Data Release 3 (DR3) provides unprecedented all-sky catalogs for eclipsing, spectroscopic, and astrometric binaries \citep{2023A&A674A1G}. 
The first {\it Gaia} catalogue of eclipsing-binary candidates contains 2\,184\,477 sources spanning a range of brightnesses from a few magnitudes up to 20 magnitudes in the {\it Gaia} G-band \citep{Mowlavi2023A&A674A16M}.
For each candidate, a geometric model of its G-band light curve is constructed by fitting to the G-band time series up to two Gaussian and a cosine \citep{Mowlavi2017}, and provide the frequency measurements.  
Spectroscopic binaries in {\it Gaia} DR3 include single-lined spectroscopic binaries (SB1), single-lined spectroscopic binaries with circular orbit (SB1C), 
double-lined spectroscopic binaries (SB2), and double-lined spectroscopic binaries with circular orbit (SB2C).
All of the orbital solutions and associated parameters for these binary are publicly available and tabulated 
in the Vizier tables of the {\it Gaia} DR3 Non-single stars (NSS) catalog\footnote{https://cdsarc.cds.unistra.fr/viz-bin/cat/I/357\#/browse} \citep{Gaia2022}.
In the \texttt{nss\_two\_body\_orbit} table, there are 181\,327 SB1, 202 SB1C, 4\,630 SB2 and 746 SB2C. 

The selection of targets was made by combining the {\it Gaia} DR3 eclipsing binary catalogue \citep{Mowlavi2023A&A674A16M}  with the {\it Gaia} DR3 spectroscopic binary catalogue \citep{Gaia2022}.
To build the initial sample, from which we can identify EL CVn-like stars, 
we cross-matched the 2\,184\,477 sources with the catalogue of SB1, SB1C, SB2, and SB2C 
by {\it Gaia} source ID via the TOPCAT software\footnote{https://www.star.bris.ac.uk/~mbt/topcat/}, 
acquiring 2\,407, 8, 407 and 126 targets, respectively. 
The orbital periods of EL CVn binaries typically fall within the range of 0.5 to 3 days \citep{Maxted2014, vanRoestel2018}.
The observed light curves display boxy, shallow eclipses ($\leq 0.1$ mag depth) 
with an approximate phase difference of $\sim$ 0.5, suggesting a small eccentricity.  
Based on the observed properties for the known EL CVn binaries, we apply the following criteria to the selection: (1) The period in the 0.5 - 3 days range; 
(2) The eccentricity smaller than 0.3;
(3) The orbital frequency differences between the EB and SB solutions smaller than 0.01 $d^{-1}$;
(4) The absolute value of the phase difference between two eclipse in the 0.4 - 0.6 range;
(5) The absolute value of the depth difference between two eclipse smaller than 0.15; 
(6) The semi-major axis of the primary star ($a_1\sin i$) in the 0.3 - 0.85 range. 

\begin{figure*}
      \centering
      \begin{minipage}{0.48\linewidth}
         \centering
         \includegraphics[width=\textwidth]{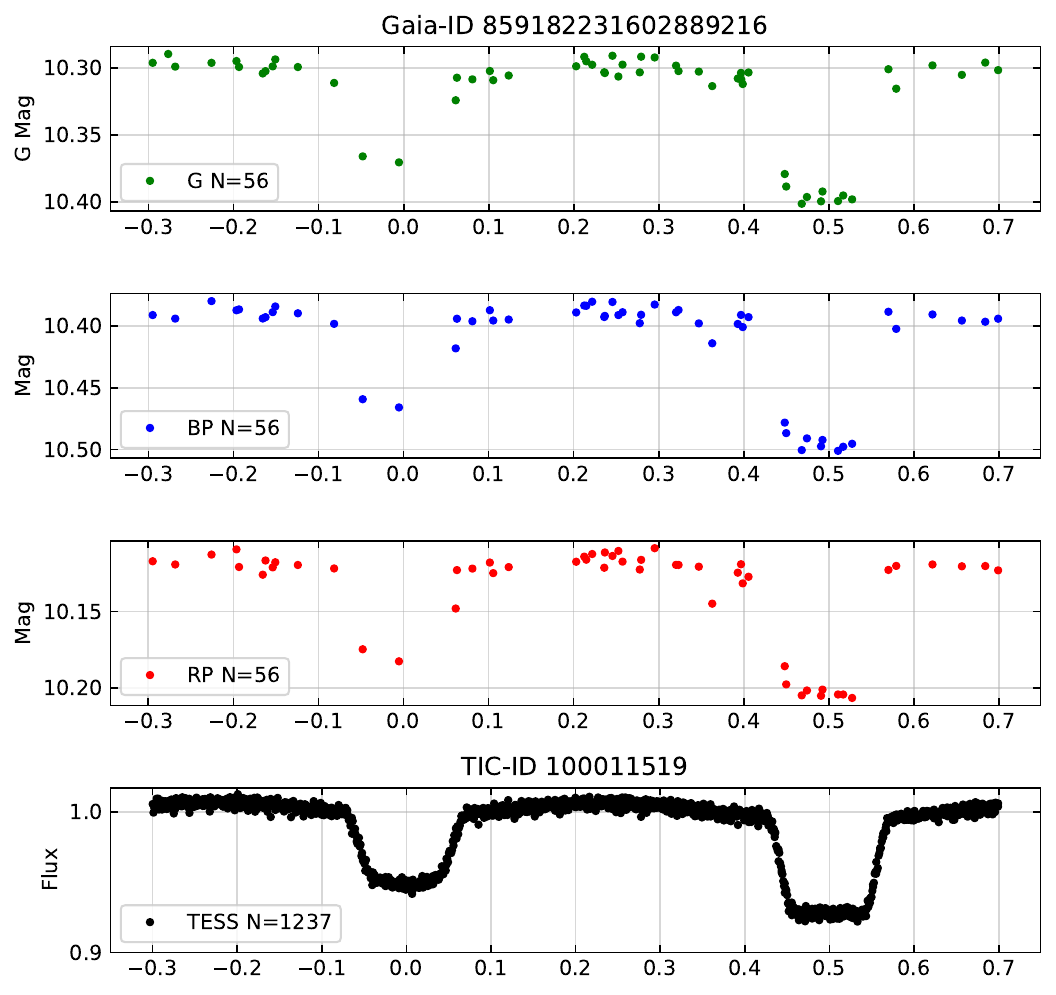}
      \end{minipage}
      \quad
      \begin{minipage}{0.48\linewidth}
         \centering
         \includegraphics[width=\textwidth]{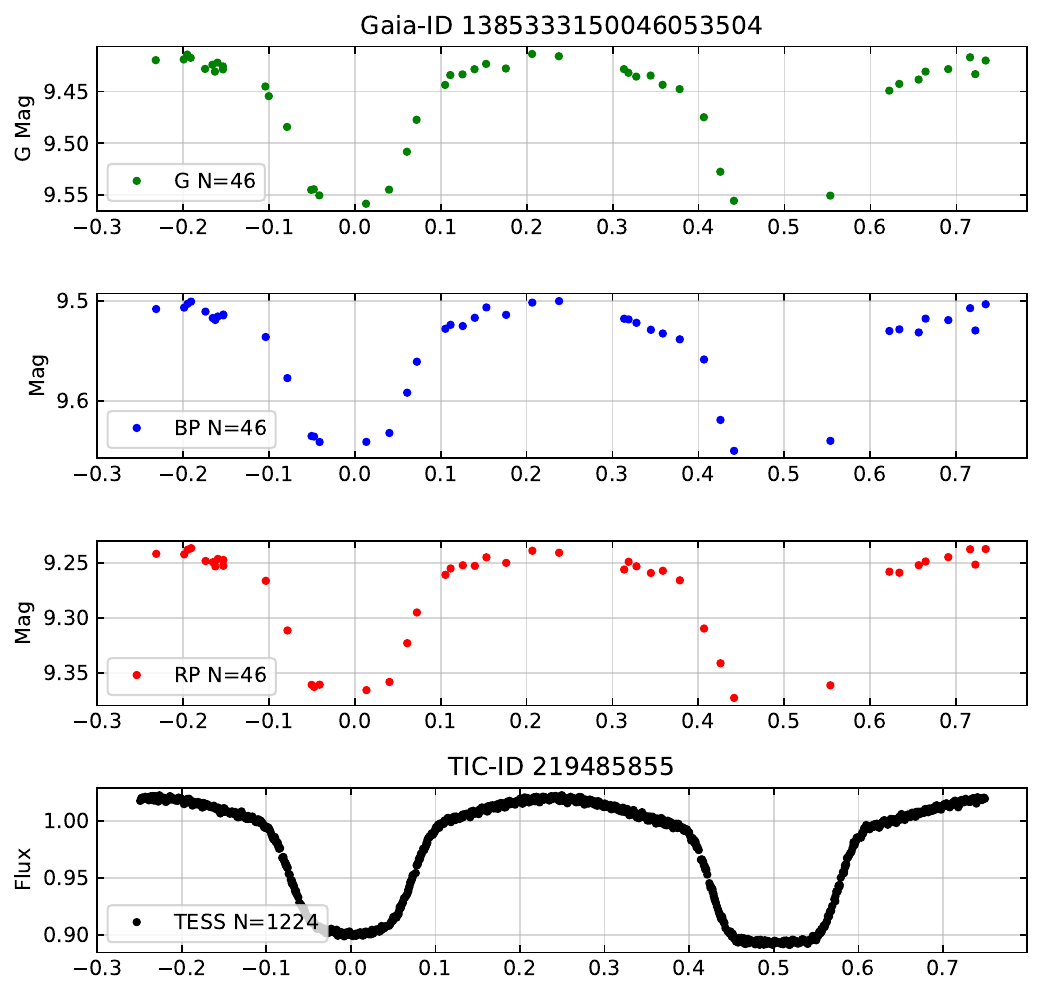}
      \end{minipage}
      \quad
      \begin{minipage}{0.48\linewidth}
         \centering
         \includegraphics[width=\textwidth]{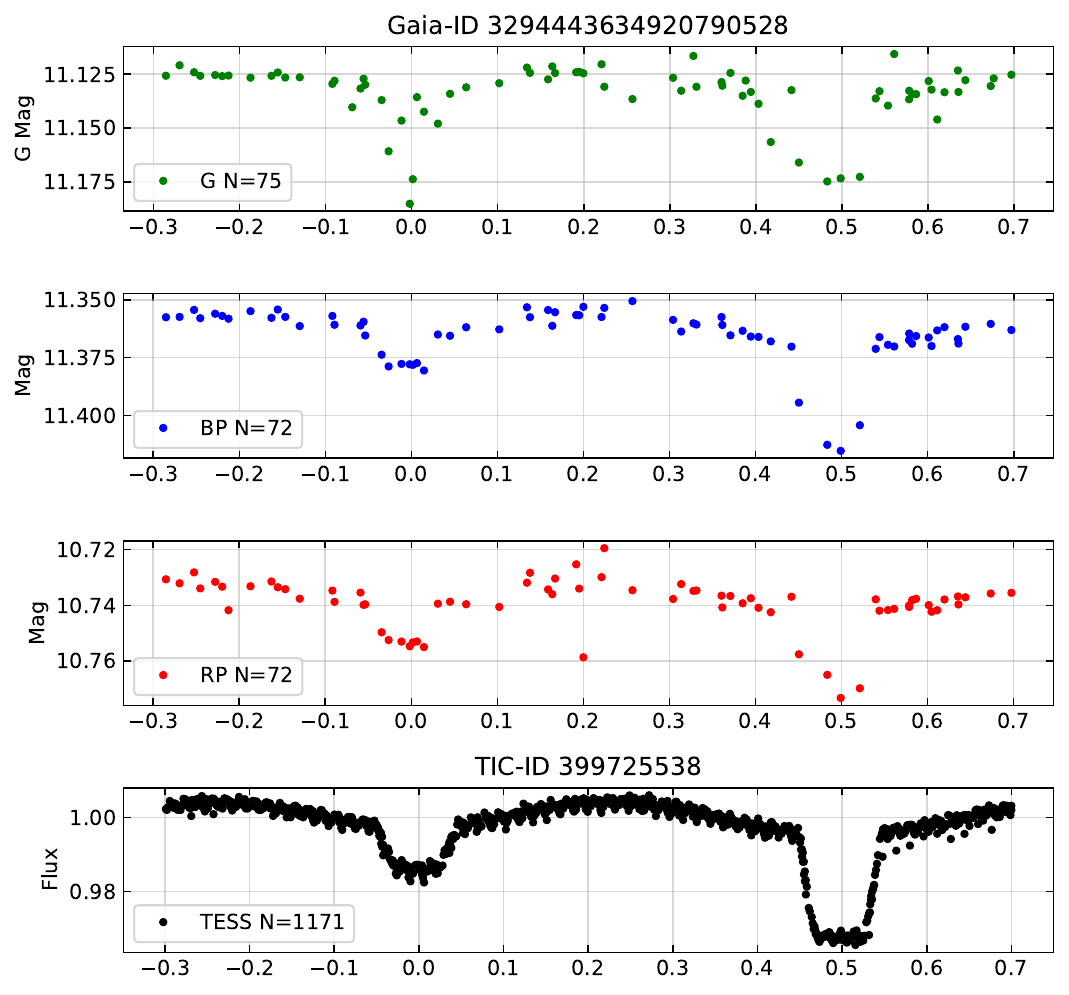}
      \end{minipage}
      \quad
      \begin{minipage}{0.48\linewidth}
         \centering
         \includegraphics[width=\textwidth]{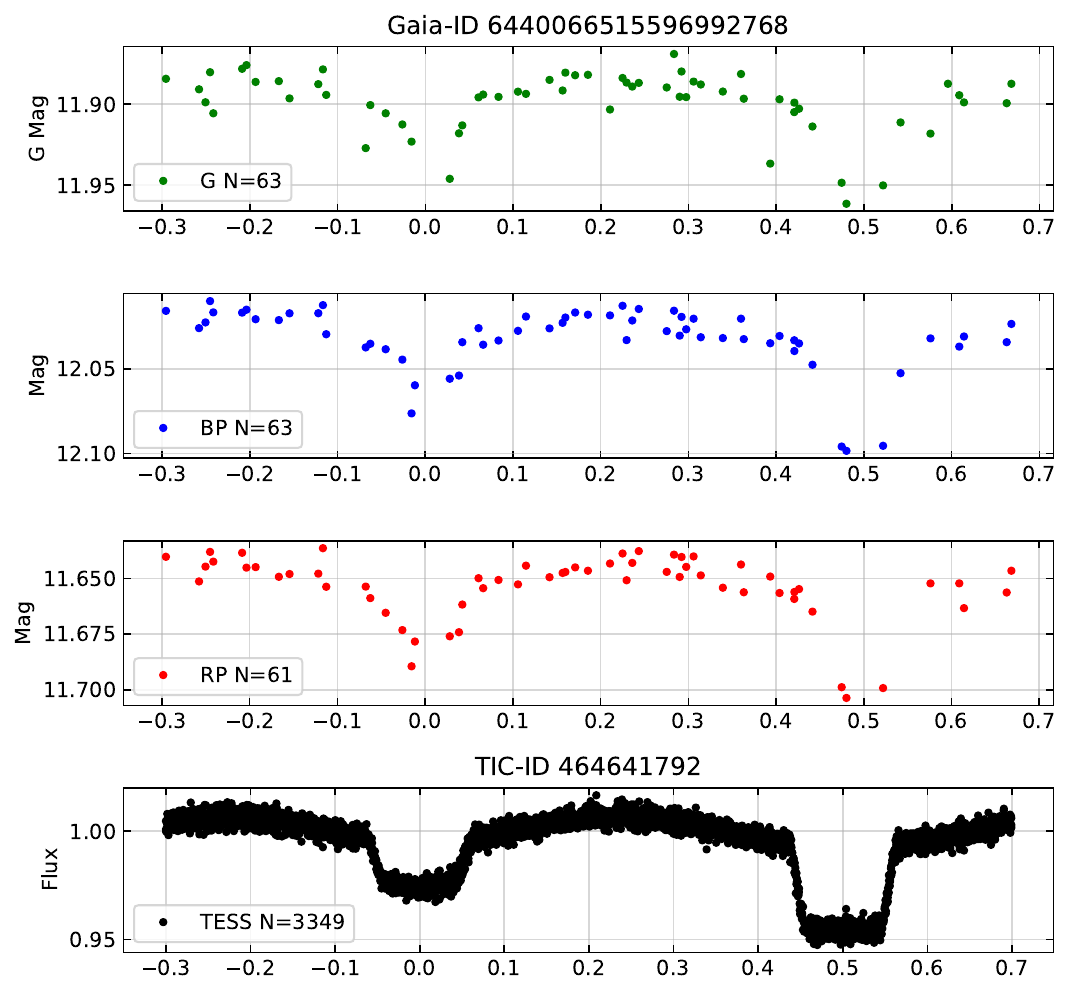}
      \end{minipage}
   \caption{{\it Gaia} and {\it TESS} photometry of four new EL CVn-type binaries.}
   \vspace{-1em}   
      \label{fig1}
\end{figure*}

These criteria yielded results of 174, 0, 3, and 6 systems for the categories EB+SB1, EB+SB1C, EB+SB2, and EB+SB2C, respectively. 
Subsequently, we conducted a visual examination of the {\it TESS} photometry data, identifying a total of 13 EL CVn-like stars, which are outlined in Table \ref{tab1}.
Upon cross-referencing the list of known EL CVn binaries, our study confirms the existence of nine previously known samples and identifies four stars as new discoveries.   
Basic information of the 13 EL CVn stars and the corresponding {\it Gaia} NSS orbital solutions are listed in Table \ref{tab1} and Table \ref{tab2}, respectively. 
Figure \ref{fig1} shows the folded {\it Gaia} $G$, $G_{BP}$, and $G_{RP}$ photometry for the four new EL CVn-type systems, together with the {\it TESS} data. 
This study focuses on the four newly identified EL CVn-type binaries.

\setlength{\tabcolsep}{6.0pt}
\renewcommand\arraystretch{1.0}
\begin{table}
   \caption{\centering Basic information of the 13 EL CVn-type binaries}
    \begin{threeparttable}
    \begin{tabular*}{\tblwidth}{CCCCCC}
\hline
\hline
{\it Gaia} DR3 source ID           	&TIC ID      &P $(d)$\tnote{a}     &{\it Gaia} mag   &{\it TESS} mag    &Notes   \\
\hline
859182231602889216 	 &100011519	   	&1.735511             &10.3005        	&10.1669           	&new   \\
1385333150046053504     &219485855	   	&0.659999             &9.4513        	 &9.3162            	&new   \\
3294443634920790528 	&399725538 	   	&1.293181             &11.1291        	 &10.8100            	&new   \\
6440066515596992768 	&464641792	   	&1.369719             &11.9069        	 &11.7177            	&new   \\
480611242765860992      &400028476     		&0.644107            &10.4856           	 &10.2188         	& EL CVn\tnote{b} \\
2048990809445098112 	&378080617     	&1.368197            &10.9688           	 &10.7093         	& EL CVn\tnote{b} \\
1987680971620394624 	&197604137     	&1.161897            &11.8662           	 &11.5850         		& EL CVn\tnote{b} \\
6637219674994191744 	&120066508     	&1.273373            &11.6504           	 &11.3419         		& EL CVn\tnote{b}\\
5087757377681887232 	&121078334     	&0.928595            &9.5519            	 &9.3429          		& EL CVn\tnote{b,c}\\
5747619351127941504 	&54957535      		&0.792833            &10.5075           	 &10.2459         	& EL CVn\tnote{c}\\
1549628911977705856 	&165371937     	&0.795641            &9.4146            	 &9.3222          		& EL CVn\tnote{c}\\
1735190117848195712 	&466940572     	&1.563113            &11.7510           	 &11.5716         		& EL CVn\tnote{c}\\
4660987986700490368 	&149160359     	&1.120692            &10.1569           	 &9.9927          		& EL CVn\tnote{d}\\
\hline
\end{tabular*}
\end{threeparttable}
\begin{tablenotes}
    \footnotesize
    \item{$^a$} computed by 1/frequency and the frequency is provided by \cite{Mowlavi2023A&A674A16M}.
    \item{$^b$} identified as EL CVn-type candidates by \cite{Gaia2023A&A674A34G} .
    \item{$^c$} identified as EL CVn-type binaries by \cite{Maxted2014}.
    \item{$^d$} identified as EL CVn-type binaries by \cite{Wang2020}.
   %  \item{$^e$} magnitude comes from \cite{Paegert2022}.
\end{tablenotes}
\label{tab1}
\end{table}

\setlength{\tabcolsep}{6.0pt}
\renewcommand\arraystretch{1.0}
\begin{table}
   \caption{\centering The {\it Gaia} NSS model solutions for the 13 EL CVn-type binaries}
    \begin{threeparttable}
    \begin{tabular*}{\tblwidth}{CCCCCCC}
\hline
\hline
Gaia DR3 source ID    		&ecc   &e\_ecc   &K1 (km/s)   &e\_K1 (km/s)   &$a_{1}\sin i$ ($R_\odot$)    &e\_$a_{1}\sin i$ ($R_\odot$)   \\
\hline
859182231602889216    		&0.105	&0.053	&21.817	&1.088	&0.744	&0.038\\
1385333150046053504      	&0.030	&0.077	&24.063	&1.466	&0.313	&0.019\\
3294443634920790528    		&0.101	&0.063	&22.091	&1.330	&0.561	&0.034\\
6440066515596992768       	&0.193	&0.085	&28.116	&2.370	&0.743	&0.064\\
480611242765860992      		&0.086	&0.045	&28.721	&1.469	&0.364	&0.018\\
2048990809445098112    		&0.080	&0.075	&22.281	&1.672	&0.599	&0.045\\
1987680971620394624    		&0.061	&0.085	&30.415	&3.573	&0.695	&0.082\\
6637219674994191744   		&0.057	&0.092	&32.277	&1.772	&0.807	&0.045\\
5087757377681887232     	&0.049	&0.037	&24.278	&0.938	&0.445	&0.017\\
5747619351127941504   		&0.116	&0.085	&30.571	&3.296	&0.474	&0.052\\
1549628911977705856   		&0.040	&0.041	&30.430	&1.257	&0.478	&0.020\\
1735190117848195712    		&0.140	&0.141	&19.127	&2.278	&0.578	&0.070\\
4660987986700490368     	&0.070	&0.085	&22.519	&1.965	&0.495	&0.043\\
\hline
\end{tabular*}
\end{threeparttable}
\begin{tablenotes}
   \footnotesize
   \item{Notes.} This table gives the eccentricity ($ecc$), the semi-amplitude of the velocity curve ($K_1$), the semi-major axis of the primary star ($a_1\sin i$) and its corresponding error. The first five columns of data are from {\it Gaia} DR3 spectroscopic binaries catalogue \citep{Gaia2022}.
\end{tablenotes}
\label{tab2}
\end{table}

\section{Binary light curve modeling}
\subsection{{\it TESS} photometry}
TIC 100011519 has been observed in Sectors 14, 15, 21, 41, and 48.
TIC 219485855 has been observed in Sectors 24, 25, 50, 51, and 52.
TIC 399725538 has been observed in Sectors 5 and 32.
TIC 464641792 has been observed in Sectors 13, 27, 66, and 67.
The {\it TESS} light-curve (LC) files, which were produced by the {\it TESS}
Science Processing Operations Center \citep{Jenkins2016}, were downloaded using the Lightkurve software\footnote{http://docs.Lightkurve.org} \citep{LightkurveCollaboration2018}.
We utilized the Simple Aperture Photometry (SAP\_FLUX) data instead of the Pre-search Data Conditioned Simple Aperture Photometry (PDCSAP\_FLUX), 
as the latter is optimized for planet transits and may have negative impacts on eclipsing binary data \citep{Slawson2011}.
We individually detrended the raw light curve of each 13.7 day orbit of the {\it TESS} spacecraft following the procedure of  \cite{Wang2020}.
Finally, we got 1\,237, 1\,224, 1\,171, and 3\,349 valid datas for TIC 100011519, TIC 219485855, TIC 399725538, and TIC 464641792, respectively.
The phased light curves for the four EL CVn stars are displayed in the Figure \ref{fig2}.

\subsection{Spectroscopic orbital elements}
According to the {\it Gaia} DR3 spectroscopic binary catalogue \citep{Gaia2022}, the four EL CVn systems are single-lined spectroscopic binaries.
The orbital elements of the four binary systems, including the orbital eccentricity ($e_{orb}$), the semi-amplitude of the velocity curve ($K_1$), 
and the orbital period ($P_{orb}$), were derived from the \texttt{nss\_two\_body\_orbit} table.
With the three accessible parameters, we calculated the semi-major axis of the primary component star ($a_1\sin i$) 
by applying the formalism of $a_{1}\sin i$ =  $\frac {\sqrt{1~ - ~e_{orb}^2}}{2\pi}$$K_{1}P_{orb}$. 
The uncertainties were estimated by using Monte Carlo error propagation.
For each star, 10\,000 random values from a Gaussian distribution were generated for the input parameters. 
Simultaneously, these random input values were used to calculate the corresponding output values. 
The standard deviation of the output values was then adopted as the uncertainties of $a_1\sin i$.   
The results are given in Table \ref{tab2}.  

\subsection{Photometric solutions}
\begin{figure*}[!t]
   % \vspace{0em}
      \centering
      \begin{minipage}{0.48\linewidth}
         \centering
         \includegraphics[width=\textwidth]{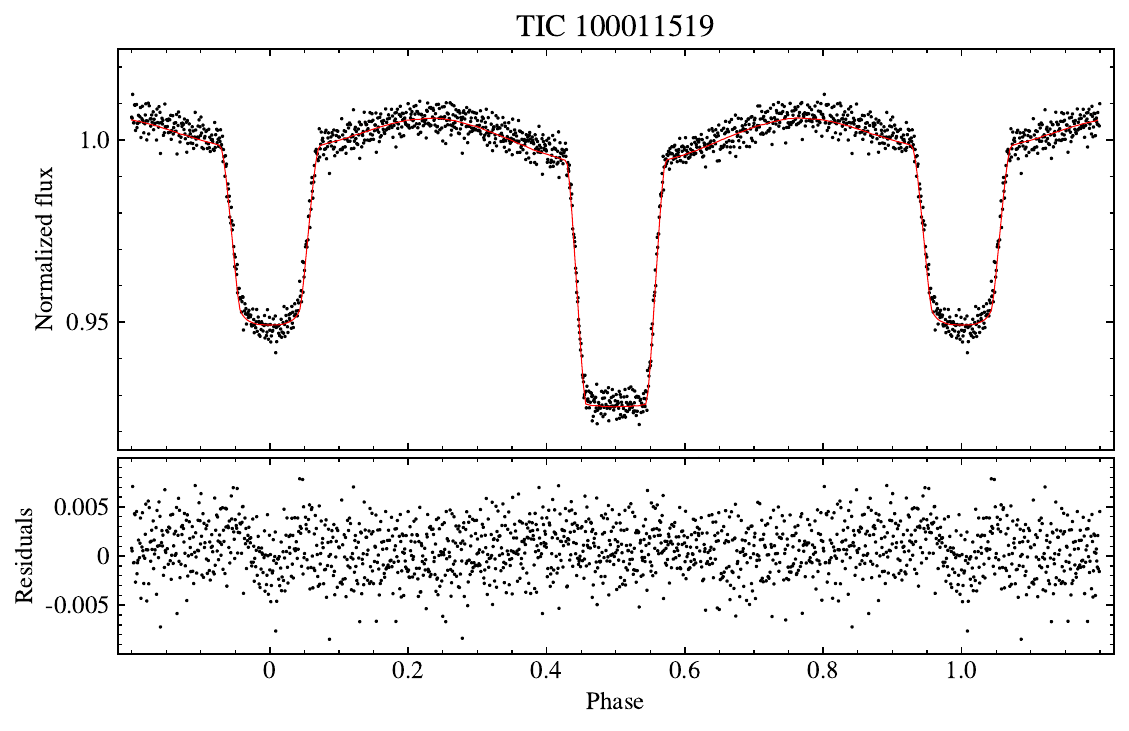}
      \end{minipage}
      % \quad
      \begin{minipage}{0.48\linewidth}
         \centering
         \includegraphics[width=\textwidth]{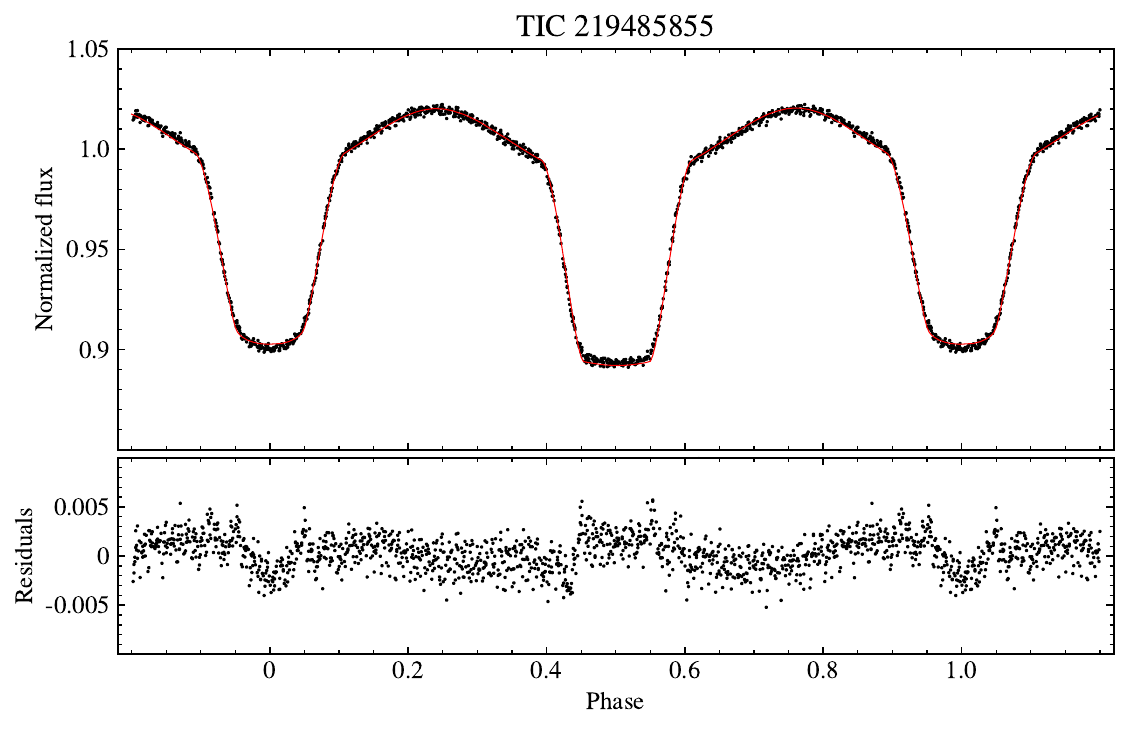}
      \end{minipage}
      \begin{minipage}{0.48\linewidth}
         \centering
         \includegraphics[width=\textwidth]{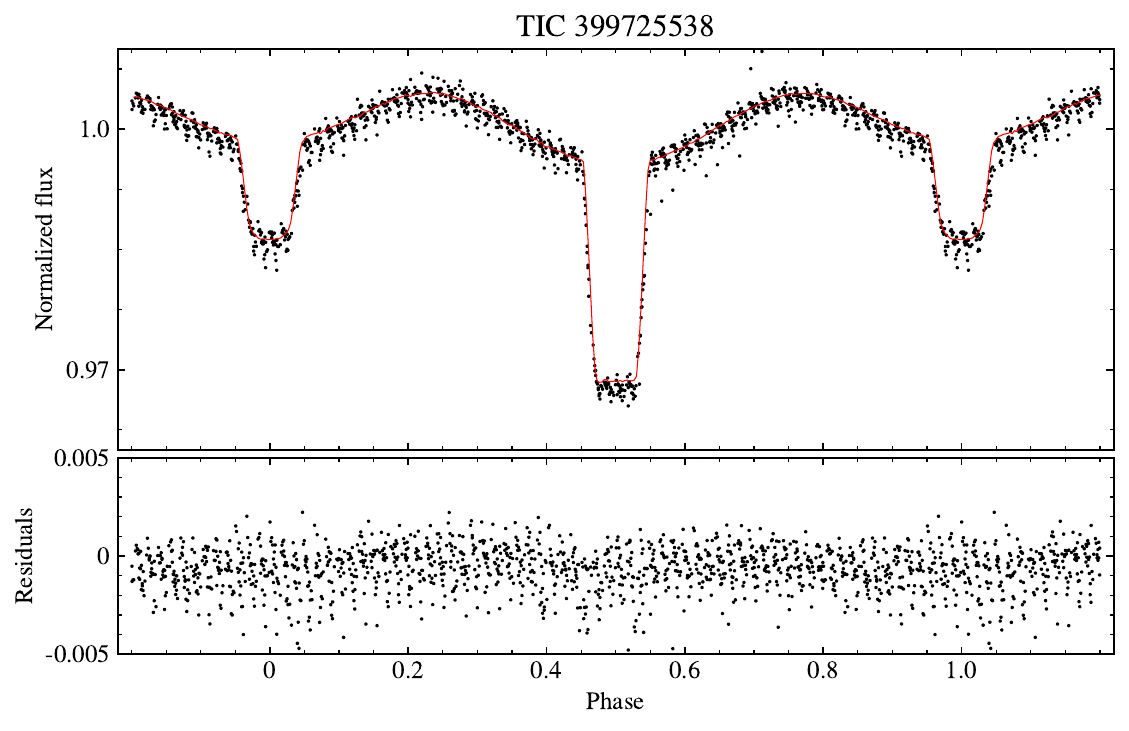}
      \end{minipage}
      \begin{minipage}{0.48\linewidth}
         \centering
         \includegraphics[width=\textwidth]{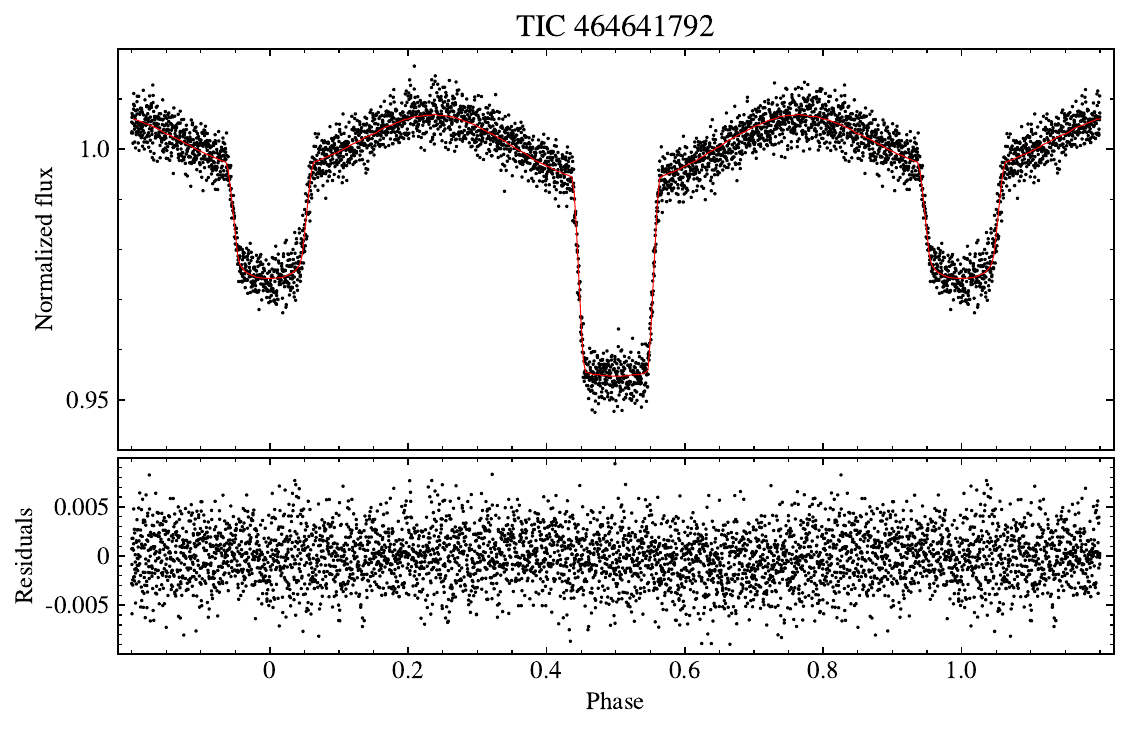}
      \end{minipage}

   \caption{Top panel: the phased light curves of the four EL CVn-type systems along with theoretical synthesis (red solid lines). 
   Bottom panel: the residuals of light curves calculated as observed minus the theoretical value.}
   \vspace{-1em}   
      \label{fig2}
\end{figure*}

We applied the Physics Of Eclipsing BinariEs software\footnote{http://www.phoebe-project.org} (PHOEBE, \citealt{Prsa2018}) 
to model the light curves and determine the photometric solutions of the four EL CVn-type systems.
In this study, we designate the massive primary star as "star 1" (subscript `1'), while the hotter but less massive star is designated as "star 2" (subscript `2').
The initial effective temperature for star 1 was obtained from \cite{Paegert2022}'s work. 
For the secondary star, an initial temperature of 10\,000 K was set, considering the temperature range of pre-He WDs, which is typically between 7\,900 and 17\,000 K \citep{vanRoestel2018}.
Since the majority of known EL CVn-type binaries exhibit a very low mass ratio of approximately 0.1, we set the initial mass ratio to be 0.1.
We utilized bolometric albedos of $A_{1}$ = $A_{2}$ =1.0, along with gravity-darkening exponents of $g_{1}$ = $g_{2}$ = 1.0.  
The free parameters in our model were: time of superior conjunction ($t_{0,supconj}$), 
the orbital period ($P_{orb}$), the mass ratio ($q_{orb}$=$M_{2}/M_{1}$), 
the semi-major axis of primary star $a_1\sin i$, eccentricity ($e_{orb}$), the orbital inclination ($i_{orb}$), 
the effective temperatures of both stars ($T_{eff,1}$, $T_{eff,2}$),
the radii of both stars ($R_{equiv,1}$, $R_{equiv,2}$), and the passband luminosity of star 1 ($L_{pb,lc01}$).
We employed the emcee sampler, incorporated in PHOEBE, to investigate the parameter spaces, 
determine the optimal solution, and evaluate the uncertainties \citep{Foreman2013, Foreman2019}. 
For this analysis, we employed 40 walkers, each with a chain length of 1500. 
To assess convergence, we visually inspected the ln(probability) plot, which illustrates the ln(probability) as a function of iteration.
After approximately 250 iterations, it was observed that all walkers for the four binary systems reached a stationary state when examining the ln(probability) plot.
To ensure convergence, the ``burn-in" step for the four binary systems was set to 500.
The fitted parameters for the best-fit model are presented in Table \ref{tab3}.
The synthetic light curve resulting from the final PHOEBE binary model is shown as solid line in the top panel of Figure \ref{fig2}.
The corresponding residuals, calculated by original light curve minus the final binary model, are displayed in the lower panels of each subfigure.  
We display the parameter posterior distributions and their interdependencies for the PHOEBE model in Figures 5-8.

\setlength{\tabcolsep}{6pt}
\renewcommand\arraystretch{1.5}
\begin{table}
\caption{\centering Photometric solutions and physical parameters of the four binary systems}
\begin{tabular*}{\tblwidth}{@{}LRRRR@{}}
    \label{tab3}\\
\hline
\hline
Parameters      				&TIC 100011519       					&TIC 219485855       				&TIC 399725538       					&TIC 464641792\\
\hline
$t_{0,supconj}$ (days)         	&$1601.5978_{-0.0037}^{+0.0041}$        	&$1900.0945_{-0.0043}^{+0.0043}$                   &$1400.9777_{-0.0022}^{+0.0018}$           &$2000.4595_{-0.0019}^{+0.0021}$  \\
$P_{orb}$ (days)               	  	&$1.735248_{-0.000074}^{+0.000066}$          		&$0.660002_{-0.000043}^{+0.000039}$         	  &$1.293273_{-0.000043}^{+0.000054}$             &$1.369715_{-0.000057}^{+0.000049}$        \\
$q=M_{2}/M_{1}$                	 &$0.0979_{-0.0070}^{+0.0065}$              		&$0.099_{-0.019}^{+0.018}$               	  &$0.0772_{-0.0038}^{+0.0054}$            	     &$0.1206_{-0.0087}^{+0.0075}$               \\         
$i$ (deg)                      	 	 &$84.3_{-1.2}^{+1.5}$                  		&$80.3_{-1.7}^{+2.0}$                    		  &$75.63_{-0.66}^{+0.69}$                 	     &$84.3_{-1.6}^{+2.3}$ 		            \\                 
$a_{1}\sin i$ ($R_{\odot}$)   	 &$0.66_{-0.046}^{+0.055}$                  		&$0.281_{-0.053}^{+0.054}$            	  &$0.435_{-0.033}^{+0.029}$               	     &$0.668_{-0.062}^{+0.057}$			 \\
$T_{eff,1}$ (K)                		&$8018.0_{-59.0}^{+57.0}$               		&$7797.0_{-78.0}^{+76.0}$                	  &$7398.0_{-83.0}^{+83.0}$              	      &$7516.0_{-67.0}^{+70.0}$		 \\
$T_{eff,2}$ (K)                		&$9452.0_{-84.0}^{+78.0}$               		&$7901.0_{-81.0}^{+97.0}$                	  &$10740.0_{-110.0}^{+120.0}$                 	      &$9943.0_{-90.0}^{+92.0}$  		\\
$e_{orb}$                      		&$0.00065_{-0.00043}^{+0.00066}$          		&$0.0021_{-0.0014}^{+0.0021}$         	   &$0.00093_{-0.00057}^{+0.00084}$             &$0.00093_{-0.00062}^{+0.00090}$ \\
$L_{pb,lc01}$ (W)              	&$11.948_{-0.011}^{+0.012}$                 		&$11.920_{-0.049}^{+0.051}$              	   &$12.391_{-0.0053}^{+0.0058}$           	     &$12.2937_{-0.0121}^{+0.0096}$   \\
\hline
\multicolumn{5}{c}{Stellar parameters derived by the PHOEBE code}\\
\hline
$a$ ($R_{\odot}$)              	&$7.49_{-0.28}^{+0.26}$                   			&$3.19_{-0.18}^{+0.20}$                  		   &$6.2_{-0.25}^{+0.26}$                  		&$6.24_{-0.28}^{+0.28}$  \\ 
$M_1$ ($M_{\odot}$)            	&$1.71_{-0.18}^{+0.18}$                   			&$0.91_{-0.14}^{+0.18}$                  		   &$1.77_{-0.20}^{+0.23}$                  		&$1.55_{-0.20}^{+0.22}$ \\ 
$M_2$ ($M_{\odot}$)            	&$0.166_{-0.020}^{+0.022}$             		&$0.089_{-0.021}^{+0.025}$               	   &$0.139_{-0.019}^{+0.019}$               		&$0.186_{-0.027}^{+0.033}$  \\ 
$R_1$ ($R_{\odot}$)            	&$2.710_{-0.094}^{+0.088}$                		&$1.534_{-0.082}^{+0.105}$                  		   &$2.143_{-0.077}^{+0.105}$               		&$2.159_{-0.091}^{+0.090}$  \\
$R_2$ ($R_{\odot}$)            	&$0.593_{-0.020}^{+0.023}$                		&$0.463_{-0.028}^{+0.031}$               	   &$0.237_{-0.010}^{+0.013}$               		&$0.312_{-0.014}^{+0.015}$ \\
$log$ $g_{1}$                  		&$3.805_{-0.022}^{+0.022}$                		&$4.025_{-0.031}^{+0.032}$               	   &$4.024_{-0.029}^{+0.023}$               		&$3.959_{-0.029}^{+0.032}$    \\                                         
$log$ $g_{2}$                  		&$4.110_{-0.043}^{+0.048}$                		&$4.060_{-0.12}^{+0.10}$               	   &$4.827_{-0.056}^{+0.046}$               		&$4.725_{-0.066}^{+0.051}$    \\                                         
\hline
\end{tabular*}
\end{table}

\section{Pulsational Characteristics}
\begin{figure*}
   % \vspace{0em}
      \centering
      \begin{minipage}{0.85\linewidth}
         \centering
         \includegraphics[width=\textwidth]{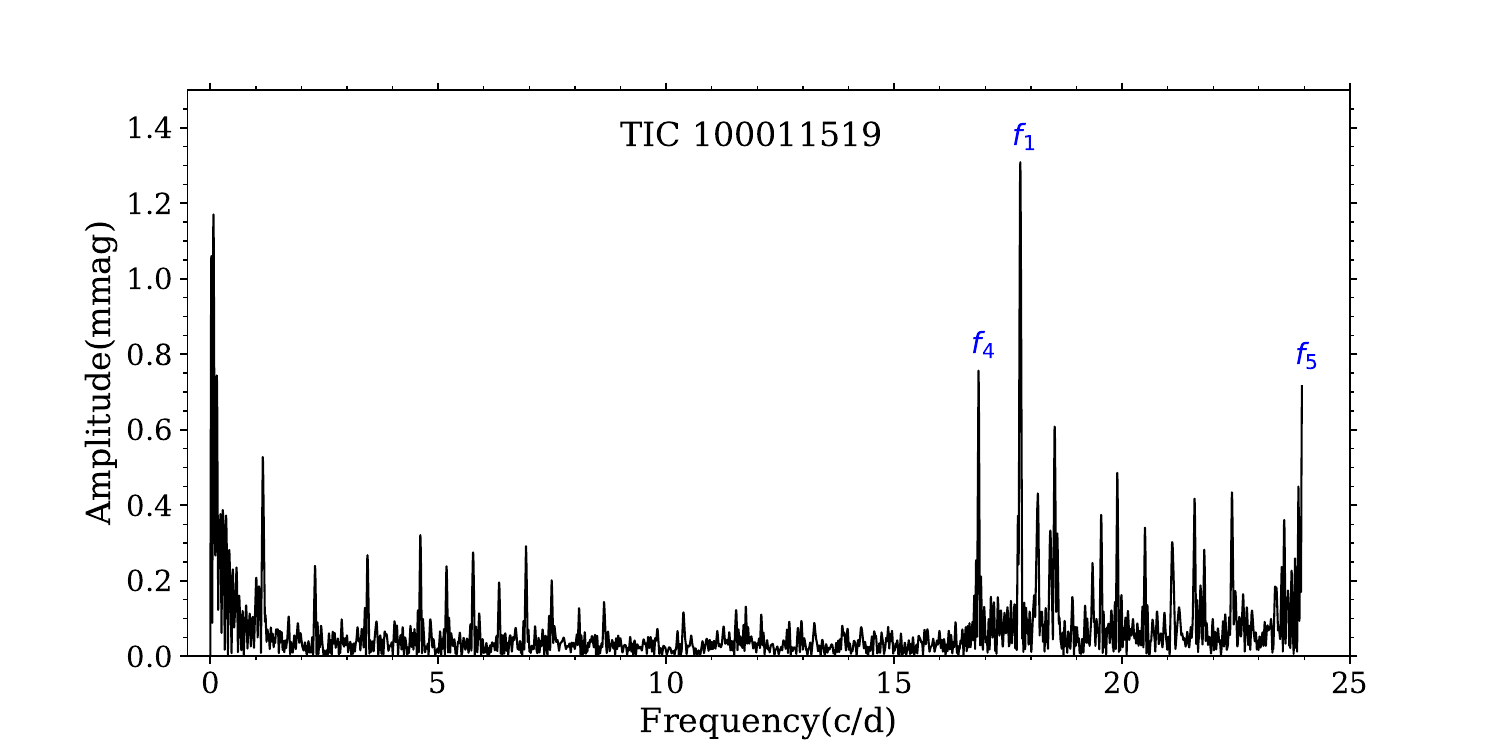}
      \end{minipage}
      % \quad
      \begin{minipage}{0.85\linewidth}
         \centering
         \includegraphics[width=\textwidth]{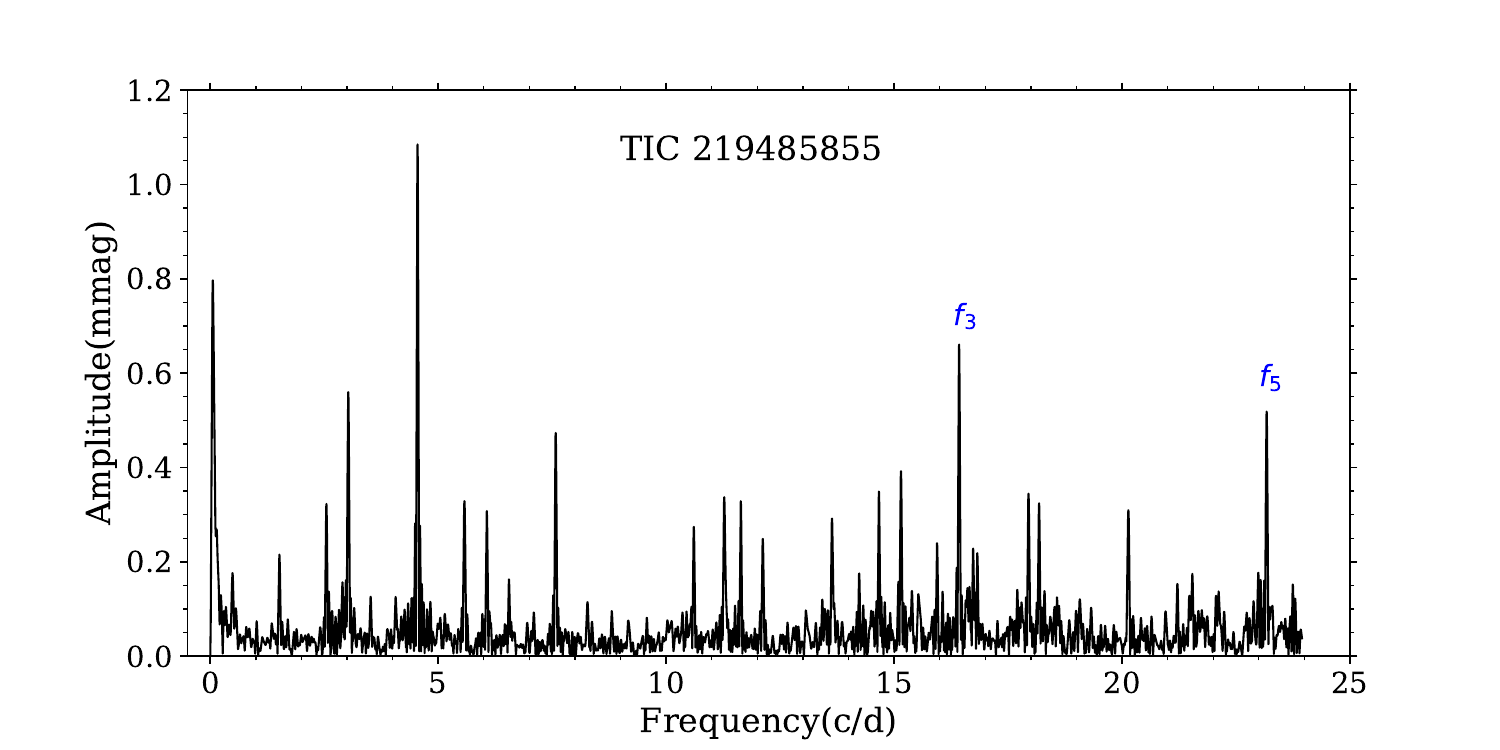}
      \end{minipage}
      \begin{minipage}{0.85\linewidth}
         \centering
         \includegraphics[width=\textwidth]{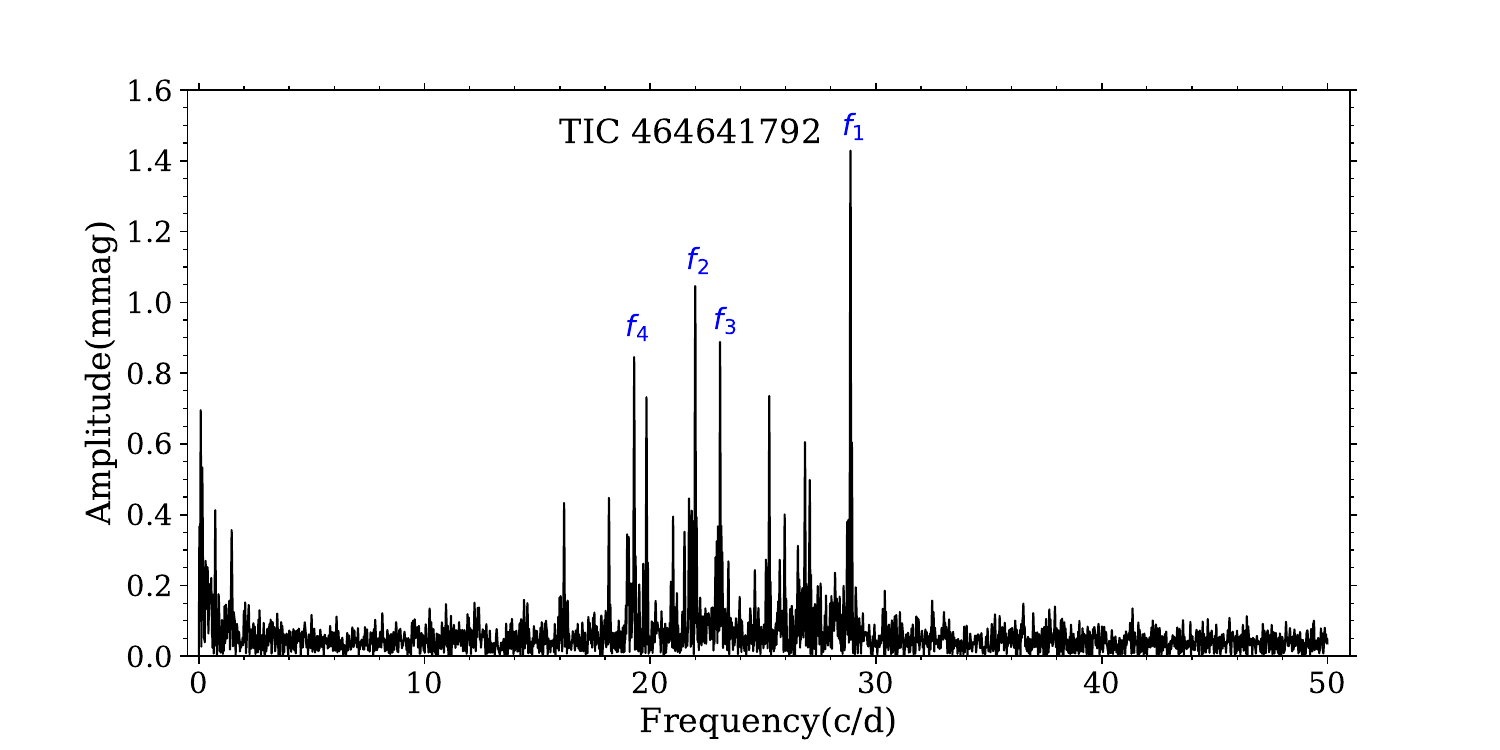}
      \end{minipage}      
   \caption{Amplitude spectra of the light residuals with the independent frequencies marked}
   \vspace{-1em}   
      \label{fig3}
\end{figure*}

In order to examine the characteristics of pulsation, a frequency analysis was performed on the residual light curves of the four binaries.
The software package Period04 \footnote{http://period04.net} \citep{Lenz2005}, 
which utilizes the discrete Fourier transform method, was employed for this analysis. 
The analysis involved an iterative prewhitening process that followed the same procedure outlined in our previous studies \citep{Chen2023NewA,Wang2023AJ}.
Only peaks with a signal-to-noise ratio (S/N) greater than 5.0, as \cite{Baran2021}'s suggestion, 
were considered for further analysis. Additionally, Period04 was used to calculate the uncertainties for all frequencies.
The results of our analysis revealed 7 significant frequencies ( S/N $\geq$ 5) for TIC 100011519, 21 frequencies for TIC 219485855 and 14 frequencies for TIC 464641792.
The detailed results of the frequency analysis can be found in Table \ref{tab4}, Table \ref{tab5} and Table \ref{tab6}.
We propose that the presence of these low frequencies below 0.1 day$^{-1}$ may be attributed to aliases or artifacts resulting from systematic trends.
Based on the detected frequencies, we conducted a search to identify potential orbital frequency harmonics ($f_{i}=kf_{orb}$) 
and linear combination frequencies ($f_{i}=f_{j}+mf_{orb}$ or $f_{i}=mf_{j}+nf_{k}$), which are listed in the last column of Table 4-6.
Figure \ref{fig3} displays the amplitude spectra of the residual light curves, with the independent frequencies marked.
For TIC 399725538, all the extracted frequencies, except for one orbital frequency harmonics, are low-frequency signals below 0.1 day$^{-1}$.
This implies that TIC 399725538 does not exhibit any intrinsic pulsations.

\setlength{\tabcolsep}{6pt}
\renewcommand\arraystretch{1.5}
\begin{table}
\caption{\centering Frequencies detected for TIC 100011519}
\begin{tabular*}{\tblwidth}{@{}CCCCCC@{}}
    \label{tab4}\\
\hline
\hline
ID & Frequency  & Amplitude  &   Phase  & S/N   & Remark \\
   & ($day^{-1}$) &  (mmag)   &    (rad/2$\pi$)   &        \\           
\hline
$f_{1}$	&17.7676$\pm$0.0014		&1.3086$\pm$0.0864	&0.6742$\pm$0.0109	&18.9	   &...\\
$f_{2}$	&0.0518$\pm$0.0050		&1.4287$\pm$0.5119	&0.9365$\pm$0.0162	&18.6          &aliases or artifacts\\
$f_{3}$	&0.0034$\pm$0.0017		&71.5618$\pm$0.0542	&0.4533$\pm$0.0401	&923.0	   &aliases or artifacts\\
$f_{4}$	&16.8506$\pm$0.0065		&0.7460$\pm$0.1072	&0.5980$\pm$0.0460	&13.3           &...\\
$f_{5}$	&23.9425$\pm$0.0024  	&0.8235$\pm$0.0824	&0.9894$\pm$0.0183	&6.1             &...\\
$f_{6}$	&18.5223$\pm$0.0029		&0.6192$\pm$0.0736	   &0.8332$\pm$0.0221	&8.2	             &$3f_{5}-3f_{1}$\\
$f_{7}$	&1.1510$\pm$0.0036		&0.5741$\pm$0.0803	   &0.1169$\pm$0.0255	&8.3	             &$2f_{orb}$\\
\hline
\end{tabular*}
\end{table}

\setlength{\tabcolsep}{6pt}
\renewcommand\arraystretch{1.5}
\begin{table}
\caption{\centering Frequencies detected for TIC 219485855}
\begin{tabular*}{\tblwidth}{@{}CCCCCC@{}}
    \label{tab5}\\
\hline
\hline
ID & Frequency  & Amplitude  &   Phase  & S/N   & Remark \\
   & ($day^{-1}$) &  (mmag)   &    (rad/2$\pi$)   &        \\           
\hline
$f_{1}$	&4.5459$\pm$0.0010	&1.0895$\pm$0.0528	&0.9626$\pm$0.0084	&37.4    &$3f_{orb}$\\
$f_{2}$	&0.0552$\pm$0.9158	&0.9522$\pm$0.5582	&0.1246$\pm$0.2640	&22.2    &aliases or artifacts\\
$f_{3}$	&16.4253$\pm$0.7954	&0.6550$\pm$0.4633	&0.2345$\pm$0.4328	&14.8    &...\\
$f_{4}$	&3.0277$\pm$2.4395	&0.5652$\pm$0.4239	&0.5196$\pm$0.3928	&18.0    &2$f_{orb}$\\
$f_{5}$	&23.1718$\pm$0.0023	&0.5235$\pm$0.0526	&0.2185$\pm$0.0171	&11.1	   &...\\
$f_{6}$	&7.5761$\pm$0.0028	&0.4715$\pm$0.0578	&0.1832$\pm$0.0195	&16.2    &$5f_{orb}$\\
$f_{7}$	&0.1075$\pm$0.7638	&0.4145$\pm$0.4603	&0.2463$\pm$0.6277	&9.7	&$2f_{2}$\\
$f_{8}$	&15.1517$\pm$1.6264	&0.3972$\pm$0.2673	&0.1056$\pm$0.4606	&9.2	&$10f_{orb}$\\
$f_{9}$	&14.6654$\pm$0.0041	&0.3411$\pm$0.0592	&0.5973$\pm$0.0309	&8.2	&$2f_{3}-12f_{orb}$\\
$f_{10}$	&17.9485$\pm$1.1925	&0.3404$\pm$0.2186	&0.2107$\pm$0.3066	&7.9	&$f_{3}+f_{orb}$\\
$f_{11}$	&11.2734$\pm$0.0036	&0.3224$\pm$0.0645	&0.1121$\pm$0.0280	&9.4	&$f_{1}-f_{3}+f_{5}$\\
$f_{12}$	&5.5755$\pm$7.5482	&0.3306$\pm$0.2226	&0.8075$\pm$0.4813	&11.5	&$2f_{3}-18f_{orb}$\\
$f_{13}$	&2.5451$\pm$0.0780	&0.3336$\pm$0.1002	&0.8055$\pm$0.0431	&9.3	&$2f_{3}-20f_{orb}$\\
$f_{14}$	&18.1790$\pm$0.8833	&0.3242$\pm$0.1660	&0.0002$\pm$0.1758	&7.7	&$12f_{orb}$\\
$f_{15}$	&20.1381$\pm$0.0041	&0.3120$\pm$0.0622	&0.2658$\pm$0.0311	&7.1	&$f_{5}-2f_{orb}$\\
$f_{16}$	&11.6363$\pm$0.8305	&0.3075$\pm$0.2097	&0.1470$\pm$0.4583	&9.0	&$2f_{3}-14f_{orb}$\\
$f_{17}$	&6.0643$\pm$0.0039	&0.3070$\pm$0.0672	&0.7887$\pm$0.0343	&10.8	&$4f_{orb}$\\
$f_{18}$	&13.6372$\pm$0.3762	&0.2894$\pm$0.0795	&0.9590$\pm$0.1253	&7.6	&$9f_{orb}$\\
$f_{19}$	&10.6061$\pm$0.0051	&0.2691$\pm$0.0574	&0.2541$\pm$0.0356	&8.3	&$7f_{orb}$\\
$f_{20}$	&12.1202$\pm$2.3451	&0.2526$\pm$0.1420	&0.2923$\pm$0.2425	&6.8	&$8f_{orb}$\\
$f_{21}$	&15.9396$\pm$10.2300	&0.2440$\pm$0.2031	&0.8515$\pm$0.4222	&5.5	&$f_{1}+f_{7}+f_{11}$\\
\hline
\end{tabular*}
\end{table}

\setlength{\tabcolsep}{6pt}
\renewcommand\arraystretch{1.5}
\begin{table}
\caption{\centering Frequencies detected for TIC 464641792}
\begin{tabular*}{\tblwidth}{@{}CCCCCC@{}}
    \label{tab6}\\
\hline
\hline
ID & Frequency  & Amplitude  &   Phase  & S/N   & Remark \\
   & ($day^{-1}$) &  (mmag)   &    (rad/2$\pi$)   &        \\           
\hline
$f_{1}$	&28.8705$\pm$1.8241	&1.4256$\pm$0.7913	&0.7584$\pm$0.2862	&18.4    &...\\
$f_{2}$	&21.9930$\pm$0.2770	&1.0370$\pm$0.3583	&0.2479$\pm$0.1390	&13.2    &...\\
$f_{3}$	&23.0929$\pm$1.0924	&0.8753$\pm$0.4878	&0.2899$\pm$0.3263	&11.2    &...\\
$f_{4}$	&19.2859$\pm$0.0014	&0.8668$\pm$0.0601	&0.2661$\pm$0.0102	&13.0    &...\\
$f_{5}$	&19.8327$\pm$0.0017	&0.7523$\pm$0.0549	&0.3276$\pm$0.0130	&10.2	   &$3f_{2}-2f_{3}$\\
$f_{6}$	&25.2706$\pm$0.0016	&0.7350$\pm$0.0625	&0.5668$\pm$0.0140	&9.2    &$f_{2}+f_{3}-f_{5}$\\
$f_{7}$	&0.0812$\pm$0.0019	&0.7922$\pm$0.0712	&0.2645$\pm$0.0121	&8.8	&$3f_{4}-2f_{1}$\\
$f_{8}$	&26.8549$\pm$0.0022	&0.5930$\pm$0.0676	&0.4972$\pm$0.0160	&7.1	&$2f_{3}-f_{4}$\\
$f_{9}$	&0.1542$\pm$0.0028	&0.5447$\pm$0.0667	&0.0149$\pm$0.0180	&6.1	&$2f_{7}$\\
$f_{10}$	&27.0613$\pm$0.0399	&0.4667$\pm$0.0804	&0.3299$\pm$0.0597	&5.6	&$f_{8}+f_{9}$\\
$f_{11}$	&16.1822$\pm$0.0609	&0.4391$\pm$0.0651	&0.0451$\pm$0.0242	&8.2	&$f_{2}-f_{1}+f_{3}$\\
$f_{12}$	&18.1697$\pm$0.7101	&0.4368$\pm$0.0916	&0.9197$\pm$0.0444	&6.9	&$f_{2}-f_{3}+f_{4}$\\
$f_{13}$	&21.7163$\pm$3.6220	&0.4234$\pm$0.1986	&0.8860$\pm$0.2822	&5.3	&$2f_{6}-f_{1}$\\
$f_{14}$	&25.9529$\pm$0.0036	&0.4066$\pm$0.0702	&0.2868$\pm$0.2698	&5.2	&$f_{2}-f_{3}+f_{10}$\\
\hline
\end{tabular*}
\end{table}

\section{Summary and Discussion}
\begin{figure*}[!t]
   % \vspace{0em}
      \centering
      \begin{minipage}{0.48\linewidth}
         \centering
         \includegraphics[width=\textwidth]{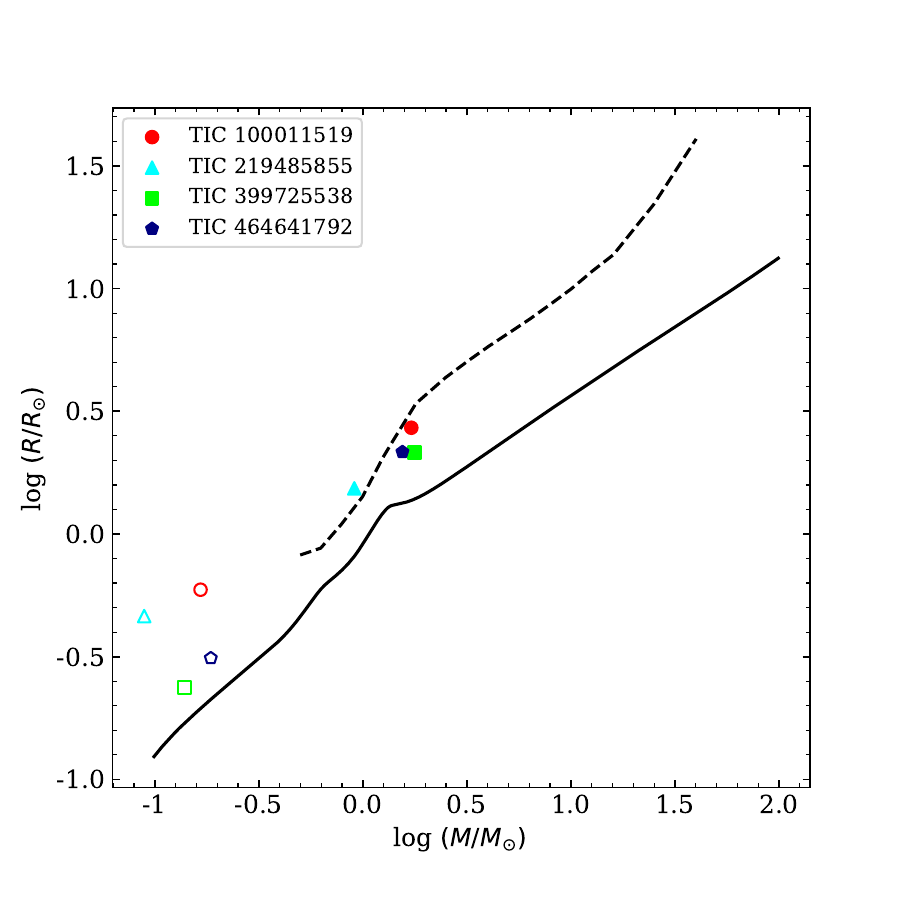}
      \end{minipage}
      % \quad
      \begin{minipage}{0.48\linewidth}
         \centering
         \includegraphics[width=\textwidth]{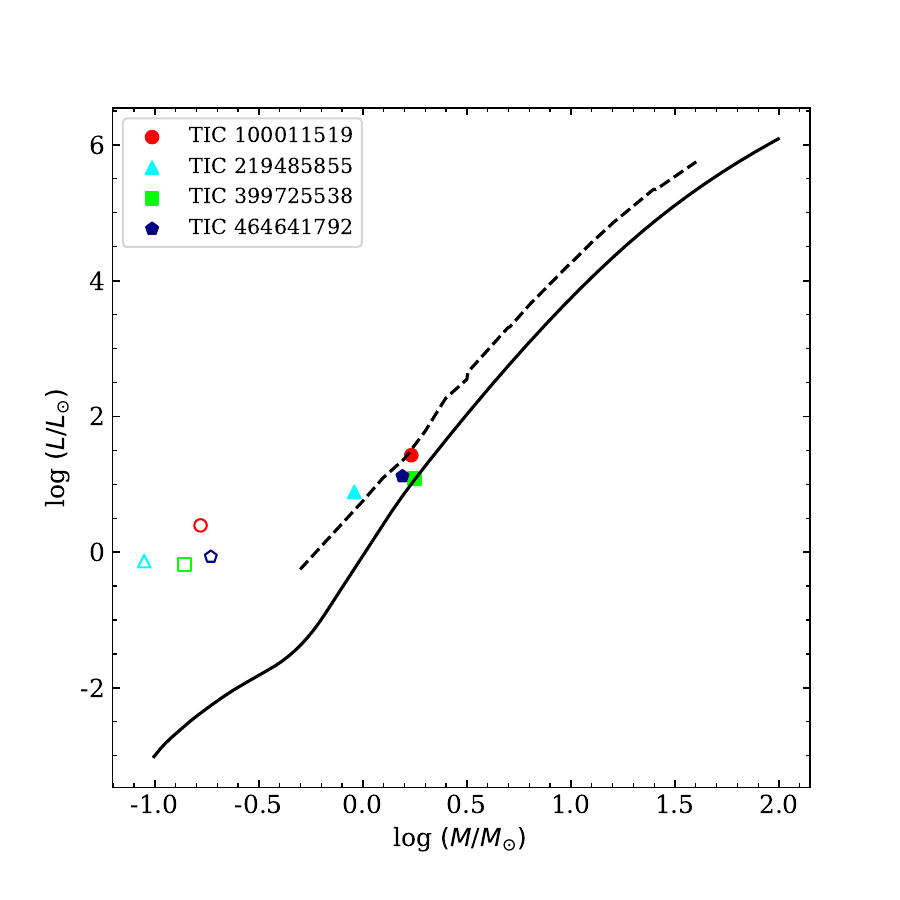}
      \end{minipage}
   \caption{Locations of the primary (filled signs) and secondary (open signs) components of the four EL-CVn type systems on the mass-radius plane (left) and mass-luminosity plane (right).
            The solid line and dashed line represent the zero-age(ZAMS) and terminal-age (TAMS) main-sequence stars, datas are taken from \cite{Pols1998}.}
   \vspace{-1em}   
      \label{fig4}
\end{figure*}

In this paper, we performed a search for EL CVn-type binaries based on the {\it Gaia} and {\it TESS} data.
Through the combination of the {\it Gaia} DR3 eclipsing binary catalogue \citep{Mowlavi2023A&A674A16M} and the {\it Gaia} DR3 spectroscopic binary catalogue \citep{Gaia2022}, 
we have identified 13 stars exhibiting EL CVn-like characteristics. 
These stars comprise of nine systems that were previously identified as EL CVn binaries and four newly discovered candidates.
The photometric solutions and absolute parameters of the four new systems were obtained by utilizing {\it TESS} photometry and spectroscopic orbital elements from \cite{Gaia2022}.
Light-curve modeling revealed very low mass ratios ($q\simeq$0.1) for all four binary systems with a detached configuration.
In Figure \ref{fig4}, We plot the four targets on the mass-radius and mass-luminosity diagrams similar to the study conducted by \cite{Ibanovglu2006}, 
the derived physical parameters presented in Table 3 suggest that the primary component of all four binary systems 
is somewhat evolved, over-luminous, and oversized A-type stars, while still falling within the main sequence band.
The less-massive components of the four binary systems are likely to be thermally
bloated ELM WDs, given their absolute physical parameters. 
Hence, our analysis suggests that the four binary systems, namely TIC 100011519, TIC 219485855, TIC 399725538, and TIC 464641792, 
are highly probable candidates for being newly discovered EL CVn-type binaries consisting of an ELM WD and an A-type star.

Multiple frequency analyses were performed on the residual light curve of the four binary systems.
The results of the frequency analysis revealed the presence of three independent pulsation frequencies for TIC 100011519, 
two frequencies for TIC 219485855, and four frequencies for TIC 464641792.
The identified independent frequencies, which range from 16.4-28.9 day$^{-1}$, are consistent with the typical frequency ranges observed in $\delta$ Sct type stars.
Additionally, the mean density of the primary component star can be derived from photometric solutions by using formalism of $\bar{\rho}$ = $M$/(4$\pi$$R$$^3$/3).
By applying the equation $Q=P_{pul}(\bar{\rho}/\bar{\rho}_{\odot})^{1/2}$, we can calculate the pulsation constant for each independent frequency. 
The calculated values for $Q$ are in the range of 0.006-0.009 days for TIC 100011519, 0.011-0.015 days for TIC 219485855, and 0.007-0.010 days for TIC 464641792.
All of the pulsation constants are found to be less than 0.033 days.
Based on these results, we propose that the primary component star in TIC 100011519, TIC 219485855 and TIC 464641792 might be a $\delta$ Sct pulsator.

The study of EL CVn binaries provides valuable insights into binary star evolution, stellar remnants, and the physical properties of compact objects. 
However, detecting ELM WDs in spectroscopic observations poses a significant challenge due to their lower brightness compared to their main-sequence (MS) companions.
To date, reliable physical parameters have only been calculated for six EL CVn systems 
using double-lined radial velocities (WASP 0247–25, \citealt{Maxted2013Natur}; KOI-81, \citealt{Matson2015ApJ};  EL CVn, \citealt{Wang2020AJ};
WASP 0131+28, \citealt{Lee2020AJ}; WASP 0843-11, \citealt{Hong2021AJ}; WASP 1625-04, \citealt{Lee2022MNRAS}).
The four newly discovered EL CVn-type binaries are sufficiently bright for high-resolution spectroscopic observations and high-cadence multicolor photometry.
We appeal to pay more attention to these stars.

\section*{Acknowledgement}
This work is supported by the National Natural Science Foundation of China (Grant No. 12373035, 12003022),
the science research grants from the China Manned Space Project,
the Sichuan Youth Science and Technology Innovation Research Team (Grant No. 21CXTD0038),
and the Innovation Research Team funds of China West Normal University (Grant No. KCXTD2022-6).

\bibliographystyle{cas-model2-names}
\bibliography{reference1}

\begin{thebibliography}{38}
\expandafter\ifx\csname natexlab\endcsname\relax\def\natexlab#1{#1}\fi
\providecommand{\url}[1]{\texttt{#1}}
\providecommand{\href}[2]{#2}
\providecommand{\path}[1]{#1}
\providecommand{\DOIprefix}{doi:}
\providecommand{\ArXivprefix}{arXiv:}
\providecommand{\URLprefix}{URL: }
\providecommand{\Pubmedprefix}{pmid:}
\providecommand{\doi}[1]{\href{http://dx.doi.org/#1}{\path{#1}}}
\providecommand{\Pubmed}[1]{\href{pmid:#1}{\path{#1}}}
\providecommand{\bibinfo}[2]{#2}
\ifx\xfnm\relax \def\xfnm[#1]{\unskip,\space#1}\fi
%Type = Article
\bibitem[{{Baran} and {Koen}(2021)}]{Baran2021}
\bibinfo{author}{{Baran}, A.S.}, \bibinfo{author}{{Koen}, C.},
  \bibinfo{year}{2021}.
\newblock \bibinfo{title}{{A Detection Threshold in the Amplitude Spectra
  Calculated from TESS Time-Series Data}}.
\newblock \bibinfo{journal}{actaa} \bibinfo{volume}{71},
  \bibinfo{pages}{113--121}.
\newblock \DOIprefix\doi{10.32023/0001-5237/71.2.3},
  \href{http://arxiv.org/abs/2106.09718}{\tt arXiv:2106.09718}.
%Type = Article
\bibitem[{{Brown} et~al.(2022){Brown}, {Kilic}, {Kosakowski} and
  {Gianninas}}]{Brown2022}
\bibinfo{author}{{Brown}, W.R.}, \bibinfo{author}{{Kilic}, M.},
  \bibinfo{author}{{Kosakowski}, A.}, \bibinfo{author}{{Gianninas}, A.},
  \bibinfo{year}{2022}.
\newblock \bibinfo{title}{{The ELM Survey. IX. A Complete Sample of Low-mass
  White Dwarf Binaries in the SDSS Footprint}}.
\newblock \bibinfo{journal}{apj} \bibinfo{volume}{933}, \bibinfo{pages}{94}.
\newblock \DOIprefix\doi{10.3847/1538-4357/ac72ac},
  \href{http://arxiv.org/abs/2207.02998}{\tt arXiv:2207.02998}.
%Type = Article
\bibitem[{{Chen} et~al.(2021){Chen}, {Tauris}, {Han} and {Chen}}]{Chen2021}
\bibinfo{author}{{Chen}, H.L.}, \bibinfo{author}{{Tauris}, T.M.},
  \bibinfo{author}{{Han}, Z.}, \bibinfo{author}{{Chen}, X.},
  \bibinfo{year}{2021}.
\newblock \bibinfo{title}{{Formation of millisecond pulsars with helium white
  dwarfs, ultra-compact X-ray binaries, and gravitational wave sources}}.
\newblock \bibinfo{journal}{mnras} \bibinfo{volume}{503},
  \bibinfo{pages}{3540--3551}.
\newblock \DOIprefix\doi{10.1093/mnras/stab670},
  \href{http://arxiv.org/abs/2103.02931}{\tt arXiv:2103.02931}.
%Type = Article
\bibitem[{{Chen} et~al.(2023){Chen}, {Wang} and {Cao}}]{Chen2023NewA}
\bibinfo{author}{{Chen}, T.}, \bibinfo{author}{{Wang}, K.},
  \bibinfo{author}{{Cao}, X.}, \bibinfo{year}{2023}.
\newblock \bibinfo{title}{{Spectroscopic and photometric study of the pulsating
  eclipsing binary HIP 28271}}.
\newblock \bibinfo{journal}{na} \bibinfo{volume}{100}, \bibinfo{pages}{101975}.
\newblock \DOIprefix\doi{10.1016/j.newast.2022.101975}.
%Type = Article
\bibitem[{{Chen} et~al.(2017){Chen}, {Maxted}, {Li} and {Han}}]{Chen2017}
\bibinfo{author}{{Chen}, X.}, \bibinfo{author}{{Maxted}, P.F.L.},
  \bibinfo{author}{{Li}, J.}, \bibinfo{author}{{Han}, Z.},
  \bibinfo{year}{2017}.
\newblock \bibinfo{title}{{The Formation of EL CVn-type Binaries}}.
\newblock \bibinfo{journal}{mnras} \bibinfo{volume}{467},
  \bibinfo{pages}{1874--1889}.
\newblock \DOIprefix\doi{10.1093/mnras/stx115},
  \href{http://arxiv.org/abs/1604.01956}{\tt arXiv:1604.01956}.
%Type = Article
\bibitem[{{Foreman-Mackey} et~al.(2019){Foreman-Mackey}, {Farr}, {Sinha},
  {Archibald}, {Hogg}, {Sanders}, {Zuntz}, {Williams}, {Nelson}, {de
  Val-Borro}, {Erhardt}, {Pashchenko} and {Pla}}]{Foreman2019}
\bibinfo{author}{{Foreman-Mackey}, D.}, \bibinfo{author}{{Farr}, W.},
  \bibinfo{author}{{Sinha}, M.}, \bibinfo{author}{{Archibald}, A.},
  \bibinfo{author}{{Hogg}, D.}, \bibinfo{author}{{Sanders}, J.},
  \bibinfo{author}{{Zuntz}, J.}, \bibinfo{author}{{Williams}, P.},
  \bibinfo{author}{{Nelson}, A.}, \bibinfo{author}{{de Val-Borro}, M.},
  \bibinfo{author}{{Erhardt}, T.}, \bibinfo{author}{{Pashchenko}, I.},
  \bibinfo{author}{{Pla}, O.}, \bibinfo{year}{2019}.
\newblock \bibinfo{title}{{emcee v3: A Python ensemble sampling toolkit for
  affine-invariant MCMC}}.
\newblock \bibinfo{journal}{The Journal of Open Source Software}
  \bibinfo{volume}{4}, \bibinfo{pages}{1864}.
\newblock \DOIprefix\doi{10.21105/joss.01864},
  \href{http://arxiv.org/abs/1911.07688}{\tt arXiv:1911.07688}.
%Type = Article
\bibitem[{{Foreman-Mackey} et~al.(2013){Foreman-Mackey}, {Hogg}, {Lang} and
  {Goodman}}]{Foreman2013}
\bibinfo{author}{{Foreman-Mackey}, D.}, \bibinfo{author}{{Hogg}, D.W.},
  \bibinfo{author}{{Lang}, D.}, \bibinfo{author}{{Goodman}, J.},
  \bibinfo{year}{2013}.
\newblock \bibinfo{title}{{emcee: The MCMC Hammer}}.
\newblock \bibinfo{journal}{pasp} \bibinfo{volume}{125}, \bibinfo{pages}{306}.
\newblock \DOIprefix\doi{10.1086/670067},
  \href{http://arxiv.org/abs/1202.3665}{\tt arXiv:1202.3665}.
%Type = Article
\bibitem[{{Gaia Collaboration}(2022)}]{Gaia2022}
\bibinfo{author}{{Gaia Collaboration}}, \bibinfo{year}{2022}.
\newblock \bibinfo{title}{{VizieR Online Data Catalog: Gaia DR3 Part 3.
  Non-single stars (Gaia Collaboration, 2022)}}.
\newblock \bibinfo{journal}{VizieR Online Data Catalog} ,
  \bibinfo{pages}{I/357}.
%Type = Article
\bibitem[{{Gaia Collaboration}(2023a)}]{Gaia2023A&A674A34G}
\bibinfo{author}{{Gaia Collaboration}}, \bibinfo{year}{2023}a.
\newblock \bibinfo{title}{{Gaia Data Release 3. Stellar multiplicity, a teaser
  for the hidden treasure}}.
\newblock \bibinfo{journal}{aap} \bibinfo{volume}{674}, \bibinfo{pages}{A34}.
\newblock \DOIprefix\doi{10.1051/0004-6361/202243782},
  \href{http://arxiv.org/abs/2206.05595}{\tt arXiv:2206.05595}.
%Type = Article
\bibitem[{{Gaia Collaboration}(2023b)}]{2023A&A674A1G}
\bibinfo{author}{{Gaia Collaboration}}, \bibinfo{year}{2023}b.
\newblock \bibinfo{title}{{Gaia Data Release 3. Summary of the content and
  survey properties}}.
\newblock \bibinfo{journal}{aap} \bibinfo{volume}{674}, \bibinfo{pages}{A1}.
\newblock \DOIprefix\doi{10.1051/0004-6361/202243940},
  \href{http://arxiv.org/abs/2208.00211}{\tt arXiv:2208.00211}.
%Type = Article
\bibitem[{{Hong} et~al.(2021){Hong}, {Lee}, {Koo}, {Park}, {Rittipruk}, {Kim},
  {Kanjanasakul} and {Han}}]{Hong2021AJ}
\bibinfo{author}{{Hong}, K.}, \bibinfo{author}{{Lee}, J.W.},
  \bibinfo{author}{{Koo}, J.R.}, \bibinfo{author}{{Park}, J.H.},
  \bibinfo{author}{{Rittipruk}, P.}, \bibinfo{author}{{Kim}, H.Y.},
  \bibinfo{author}{{Kanjanasakul}, C.}, \bibinfo{author}{{Han}, C.},
  \bibinfo{year}{2021}.
\newblock \bibinfo{title}{{The Pre-He White Dwarfs in Eclipsing Binaries. II.
  WASP 0843-11}}.
\newblock \bibinfo{journal}{aj} \bibinfo{volume}{161}, \bibinfo{pages}{137}.
\newblock \DOIprefix\doi{10.3847/1538-3881/abdd39}.
%Type = Article
\bibitem[{{Ibano{\v{g}}lu} et~al.(2006){Ibano{\v{g}}lu}, {Soydugan}, {Soydugan}
  and {Dervi{\c{s}}o{\v{g}}lu}}]{Ibanovglu2006}
\bibinfo{author}{{Ibano{\v{g}}lu}, C.}, \bibinfo{author}{{Soydugan}, F.},
  \bibinfo{author}{{Soydugan}, E.}, \bibinfo{author}{{Dervi{\c{s}}o{\v{g}}lu},
  A.}, \bibinfo{year}{2006}.
\newblock \bibinfo{title}{{Angular momentum evolution of Algol binaries}}.
\newblock \bibinfo{journal}{mnras} \bibinfo{volume}{373},
  \bibinfo{pages}{435--448}.
\newblock \DOIprefix\doi{10.1111/j.1365-2966.2006.11052.x}.
%Type = Article
\bibitem[{{Istrate} et~al.(2016){Istrate}, {Marchant}, {Tauris}, {Langer},
  {Stancliffe} and {Grassitelli}}]{Istrate2016}
\bibinfo{author}{{Istrate}, A.}, \bibinfo{author}{{Marchant}, P.},
  \bibinfo{author}{{Tauris}, T.M.}, \bibinfo{author}{{Langer}, N.},
  \bibinfo{author}{{Stancliffe}, R.J.}, \bibinfo{author}{{Grassitelli}, L.},
  \bibinfo{year}{2016}.
\newblock \bibinfo{title}{{VizieR Online Data Catalog: Low-mass helium white
  dwarfs evolutionary models (Istrate+, 2016)}}.
\newblock \bibinfo{journal}{VizieR Online Data Catalog} ,
  \bibinfo{pages}{J/A+A/595/A35}.
%Type = Inproceedings
\bibitem[{{Jenkins} et~al.(2016){Jenkins}, {Twicken}, {McCauliff}, {Campbell},
  {Sanderfer}, {Lung}, {Mansouri-Samani}, {Girouard}, {Tenenbaum}, {Klaus},
  {Smith}, {Caldwell}, {Chacon}, {Henze}, {Heiges}, {Latham}, {Morgan},
  {Swade}, {Rinehart} and {Vanderspek}}]{Jenkins2016}
\bibinfo{author}{{Jenkins}, J.M.}, \bibinfo{author}{{Twicken}, J.D.},
  \bibinfo{author}{{McCauliff}, S.}, \bibinfo{author}{{Campbell}, J.},
  \bibinfo{author}{{Sanderfer}, D.}, \bibinfo{author}{{Lung}, D.},
  \bibinfo{author}{{Mansouri-Samani}, M.}, \bibinfo{author}{{Girouard}, F.},
  \bibinfo{author}{{Tenenbaum}, P.}, \bibinfo{author}{{Klaus}, T.},
  \bibinfo{author}{{Smith}, J.C.}, \bibinfo{author}{{Caldwell}, D.A.},
  \bibinfo{author}{{Chacon}, A.D.}, \bibinfo{author}{{Henze}, C.},
  \bibinfo{author}{{Heiges}, C.}, \bibinfo{author}{{Latham}, D.W.},
  \bibinfo{author}{{Morgan}, E.}, \bibinfo{author}{{Swade}, D.},
  \bibinfo{author}{{Rinehart}, S.}, \bibinfo{author}{{Vanderspek}, R.},
  \bibinfo{year}{2016}.
\newblock \bibinfo{title}{{The TESS science processing operations center}}, in:
  \bibinfo{editor}{{Chiozzi}, G.}, \bibinfo{editor}{{Guzman}, J.C.} (Eds.),
  \bibinfo{booktitle}{Software and Cyberinfrastructure for Astronomy IV}, p.
  \bibinfo{pages}{99133E}.
\newblock \DOIprefix\doi{10.1117/12.2233418}.
%Type = Article
\bibitem[{{Justham} et~al.(2009){Justham}, {Wolf}, {Podsiadlowski} and
  {Han}}]{Justham2009}
\bibinfo{author}{{Justham}, S.}, \bibinfo{author}{{Wolf}, C.},
  \bibinfo{author}{{Podsiadlowski}, P.}, \bibinfo{author}{{Han}, Z.},
  \bibinfo{year}{2009}.
\newblock \bibinfo{title}{{Type Ia supernovae and the formation of single
  low-mass white dwarfs}}.
\newblock \bibinfo{journal}{aap} \bibinfo{volume}{493},
  \bibinfo{pages}{1081--1091}.
\newblock \DOIprefix\doi{10.1051/0004-6361:200810106},
  \href{http://arxiv.org/abs/0811.2633}{\tt arXiv:0811.2633}.
%Type = Article
\bibitem[{{Lee} et~al.(2022){Lee}, {Hong} and {Park}}]{Lee2022MNRAS}
\bibinfo{author}{{Lee}, J.W.}, \bibinfo{author}{{Hong}, K.},
  \bibinfo{author}{{Park}, J.H.}, \bibinfo{year}{2022}.
\newblock \bibinfo{title}{{The pre-He white dwarfs in eclipsing binaries - III.
  WASP 1625-04}}.
\newblock \bibinfo{journal}{mnras} \bibinfo{volume}{511},
  \bibinfo{pages}{654--661}.
\newblock \DOIprefix\doi{10.1093/mnras/stac075},
  \href{http://arxiv.org/abs/2201.02780}{\tt arXiv:2201.02780}.
%Type = Article
\bibitem[{{Lee} et~al.(2020){Lee}, {Koo}, {Hong} and {Park}}]{Lee2020AJ}
\bibinfo{author}{{Lee}, J.W.}, \bibinfo{author}{{Koo}, J.R.},
  \bibinfo{author}{{Hong}, K.}, \bibinfo{author}{{Park}, J.H.},
  \bibinfo{year}{2020}.
\newblock \bibinfo{title}{{The Pre-He White Dwarfs in Eclipsing Binaries. I.
  WASP 0131+28}}.
\newblock \bibinfo{journal}{aj} \bibinfo{volume}{160}, \bibinfo{pages}{49}.
\newblock \DOIprefix\doi{10.3847/1538-3881/ab9621},
  \href{http://arxiv.org/abs/2005.10394}{\tt arXiv:2005.10394}.
%Type = Article
\bibitem[{{Lenz} and {Breger}(2005)}]{Lenz2005}
\bibinfo{author}{{Lenz}, P.}, \bibinfo{author}{{Breger}, M.},
  \bibinfo{year}{2005}.
\newblock \bibinfo{title}{{Period04 User Guide}}.
\newblock \bibinfo{journal}{Communications in Asteroseismology}
  \bibinfo{volume}{146}, \bibinfo{pages}{53--136}.
\newblock \DOIprefix\doi{10.1553/cia146s53}.
%Type = Article
\bibitem[{{Li} et~al.(2019){Li}, {Chen}, {Chen} and {Han}}]{Li2019}
\bibinfo{author}{{Li}, Z.}, \bibinfo{author}{{Chen}, X.},
  \bibinfo{author}{{Chen}, H.L.}, \bibinfo{author}{{Han}, Z.},
  \bibinfo{year}{2019}.
\newblock \bibinfo{title}{{Formation of Extremely Low-mass White Dwarfs in
  Double Degenerates}}.
\newblock \bibinfo{journal}{apj} \bibinfo{volume}{871}, \bibinfo{pages}{148}.
\newblock \DOIprefix\doi{10.3847/1538-4357/aaf9a1},
  \href{http://arxiv.org/abs/1812.07226}{\tt arXiv:1812.07226}.
%Type = Misc
\bibitem[{{Lightkurve Collaboration} et~al.(2018){Lightkurve Collaboration},
  {Cardoso}, {Hedges}, {Gully-Santiago}, {Saunders}, {Cody}, {Barclay}, {Hall},
  {Sagear}, {Turtelboom}, {Zhang}, {Tzanidakis}, {Mighell}, {Coughlin}, {Bell},
  {Berta-Thompson}, {Williams}, {Dotson} and
  {Barentsen}}]{LightkurveCollaboration2018}
\bibinfo{author}{{Lightkurve Collaboration}}, \bibinfo{author}{{Cardoso},
  J.V.d.M.}, \bibinfo{author}{{Hedges}, C.}, \bibinfo{author}{{Gully-Santiago},
  M.}, \bibinfo{author}{{Saunders}, N.}, \bibinfo{author}{{Cody}, A.M.},
  \bibinfo{author}{{Barclay}, T.}, \bibinfo{author}{{Hall}, O.},
  \bibinfo{author}{{Sagear}, S.}, \bibinfo{author}{{Turtelboom}, E.},
  \bibinfo{author}{{Zhang}, J.}, \bibinfo{author}{{Tzanidakis}, A.},
  \bibinfo{author}{{Mighell}, K.}, \bibinfo{author}{{Coughlin}, J.},
  \bibinfo{author}{{Bell}, K.}, \bibinfo{author}{{Berta-Thompson}, Z.},
  \bibinfo{author}{{Williams}, P.}, \bibinfo{author}{{Dotson}, J.},
  \bibinfo{author}{{Barentsen}, G.}, \bibinfo{year}{2018}.
\newblock \bibinfo{title}{{Lightkurve: Kepler and TESS time series analysis in
  Python}}.
\newblock \bibinfo{howpublished}{Astrophysics Source Code Library, record
  ascl:1812.013}.
\newblock \href{http://arxiv.org/abs/1812.013}{\tt arXiv:1812.013}.
%Type = Article
\bibitem[{{Mata S{\'a}nchez} et~al.(2020){Mata S{\'a}nchez}, {Istrate}, {van
  Kerkwijk}, {Breton} and {Kaplan}}]{Mata2020}
\bibinfo{author}{{Mata S{\'a}nchez}, D.}, \bibinfo{author}{{Istrate}, A.G.},
  \bibinfo{author}{{van Kerkwijk}, M.H.}, \bibinfo{author}{{Breton}, R.P.},
  \bibinfo{author}{{Kaplan}, D.L.}, \bibinfo{year}{2020}.
\newblock \bibinfo{title}{{PSR J1012+5307: a millisecond pulsar with an
  extremely low-mass white dwarf companion}}.
\newblock \bibinfo{journal}{mnras} \bibinfo{volume}{494},
  \bibinfo{pages}{4031--4042}.
\newblock \DOIprefix\doi{10.1093/mnras/staa983},
  \href{http://arxiv.org/abs/2004.02901}{\tt arXiv:2004.02901}.
%Type = Article
\bibitem[{{Matson} et~al.(2015){Matson}, {Gies}, {Guo}, {Quinn}, {Buchhave},
  {Latham}, {Howell} and {Rowe}}]{Matson2015ApJ}
\bibinfo{author}{{Matson}, R.A.}, \bibinfo{author}{{Gies}, D.R.},
  \bibinfo{author}{{Guo}, Z.}, \bibinfo{author}{{Quinn}, S.N.},
  \bibinfo{author}{{Buchhave}, L.A.}, \bibinfo{author}{{Latham}, D.W.},
  \bibinfo{author}{{Howell}, S.B.}, \bibinfo{author}{{Rowe}, J.F.},
  \bibinfo{year}{2015}.
\newblock \bibinfo{title}{{HST/COS Detection of the Spectrum of the Subdwarf
  Companion of KOI-81}}.
\newblock \bibinfo{journal}{apj} \bibinfo{volume}{806}, \bibinfo{pages}{155}.
\newblock \DOIprefix\doi{10.1088/0004-637X/806/2/155},
  \href{http://arxiv.org/abs/1505.00817}{\tt arXiv:1505.00817}.
%Type = Article
\bibitem[{{Maxted} et~al.(2011){Maxted}, {Anderson}, {Burleigh}, {Collier
  Cameron}, {Heber}, {G{\"a}nsicke}, {Geier}, {Kupfer}, {Marsh}, {Nelemans},
  {O'Toole}, {{\O}stensen}, {Smalley} and {West}}]{Maxted2011}
\bibinfo{author}{{Maxted}, P.F.L.}, \bibinfo{author}{{Anderson}, D.R.},
  \bibinfo{author}{{Burleigh}, M.R.}, \bibinfo{author}{{Collier Cameron}, A.},
  \bibinfo{author}{{Heber}, U.}, \bibinfo{author}{{G{\"a}nsicke}, B.T.},
  \bibinfo{author}{{Geier}, S.}, \bibinfo{author}{{Kupfer}, T.},
  \bibinfo{author}{{Marsh}, T.R.}, \bibinfo{author}{{Nelemans}, G.},
  \bibinfo{author}{{O'Toole}, S.J.}, \bibinfo{author}{{{\O}stensen}, R.H.},
  \bibinfo{author}{{Smalley}, B.}, \bibinfo{author}{{West}, R.G.},
  \bibinfo{year}{2011}.
\newblock \bibinfo{title}{{Discovery of a stripped red giant core in a bright
  eclipsing binary system}}.
\newblock \bibinfo{journal}{mnras} \bibinfo{volume}{418},
  \bibinfo{pages}{1156--1164}.
\newblock \DOIprefix\doi{10.1111/j.1365-2966.2011.19567.x},
  \href{http://arxiv.org/abs/1107.4986}{\tt arXiv:1107.4986}.
%Type = Article
\bibitem[{{Maxted} et~al.(2014){Maxted}, {Bloemen}, {Heber}, {Geier},
  {Wheatley}, {Marsh}, {Breedt}, {Sebastian}, {Faillace}, {Owen}, {Pulley},
  {Smith}, {Kolb}, {Haswell}, {Southworth}, {Anderson}, {Smalley}, {Collier
  Cameron}, {Hebb}, {Simpson}, {West}, {Bochinski}, {Busuttil} and
  {Hadigal}}]{Maxted2014}
\bibinfo{author}{{Maxted}, P.F.L.}, \bibinfo{author}{{Bloemen}, S.},
  \bibinfo{author}{{Heber}, U.}, \bibinfo{author}{{Geier}, S.},
  \bibinfo{author}{{Wheatley}, P.J.}, \bibinfo{author}{{Marsh}, T.R.},
  \bibinfo{author}{{Breedt}, E.}, \bibinfo{author}{{Sebastian}, D.},
  \bibinfo{author}{{Faillace}, G.}, \bibinfo{author}{{Owen}, C.},
  \bibinfo{author}{{Pulley}, D.}, \bibinfo{author}{{Smith}, D.},
  \bibinfo{author}{{Kolb}, U.}, \bibinfo{author}{{Haswell}, C.A.},
  \bibinfo{author}{{Southworth}, J.}, \bibinfo{author}{{Anderson}, D.R.},
  \bibinfo{author}{{Smalley}, B.}, \bibinfo{author}{{Collier Cameron}, A.},
  \bibinfo{author}{{Hebb}, L.}, \bibinfo{author}{{Simpson}, E.K.},
  \bibinfo{author}{{West}, R.G.}, \bibinfo{author}{{Bochinski}, J.},
  \bibinfo{author}{{Busuttil}, R.}, \bibinfo{author}{{Hadigal}, S.},
  \bibinfo{year}{2014}.
\newblock \bibinfo{title}{{EL CVn-type binaries - discovery of 17 helium white
  dwarf precursors in bright eclipsing binary star systems}}.
\newblock \bibinfo{journal}{mnras} \bibinfo{volume}{437},
  \bibinfo{pages}{1681--1697}.
\newblock \DOIprefix\doi{10.1093/mnras/stt2007},
  \href{http://arxiv.org/abs/1310.4863}{\tt arXiv:1310.4863}.
%Type = Article
\bibitem[{{Maxted} et~al.(2013){Maxted}, {Serenelli}, {Miglio}, {Marsh},
  {Heber}, {Dhillon}, {Littlefair}, {Copperwheat}, {Smalley}, {Breedt} and
  {Schaffenroth}}]{Maxted2013Natur}
\bibinfo{author}{{Maxted}, P.F.L.}, \bibinfo{author}{{Serenelli}, A.M.},
  \bibinfo{author}{{Miglio}, A.}, \bibinfo{author}{{Marsh}, T.R.},
  \bibinfo{author}{{Heber}, U.}, \bibinfo{author}{{Dhillon}, V.S.},
  \bibinfo{author}{{Littlefair}, S.}, \bibinfo{author}{{Copperwheat}, C.},
  \bibinfo{author}{{Smalley}, B.}, \bibinfo{author}{{Breedt}, E.},
  \bibinfo{author}{{Schaffenroth}, V.}, \bibinfo{year}{2013}.
\newblock \bibinfo{title}{{Multi-periodic pulsations of a stripped red-giant
  star in an eclipsing binary system}}.
\newblock \bibinfo{journal}{nat} \bibinfo{volume}{498},
  \bibinfo{pages}{463--465}.
\newblock \DOIprefix\doi{10.1038/nature12192},
  \href{http://arxiv.org/abs/1307.1654}{\tt arXiv:1307.1654}.
%Type = Article
\bibitem[{{Mowlavi} et~al.(2023){Mowlavi}, {Holl}, {Lecoeur-Ta{\"\i}bi},
  {Barblan}, {Kochoska}, {Pr{\v{s}}a}, {Mazeh}, {Rimoldini}, {Gavras},
  {Audard}, {Jevardat de Fombelle}, {Nienartowicz}, {Garc{\'\i}a-Lario} and
  {Eyer}}]{Mowlavi2023A&A674A16M}
\bibinfo{author}{{Mowlavi}, N.}, \bibinfo{author}{{Holl}, B.},
  \bibinfo{author}{{Lecoeur-Ta{\"\i}bi}, I.}, \bibinfo{author}{{Barblan}, F.},
  \bibinfo{author}{{Kochoska}, A.}, \bibinfo{author}{{Pr{\v{s}}a}, A.},
  \bibinfo{author}{{Mazeh}, T.}, \bibinfo{author}{{Rimoldini}, L.},
  \bibinfo{author}{{Gavras}, P.}, \bibinfo{author}{{Audard}, M.},
  \bibinfo{author}{{Jevardat de Fombelle}, G.},
  \bibinfo{author}{{Nienartowicz}, K.}, \bibinfo{author}{{Garc{\'\i}a-Lario},
  P.}, \bibinfo{author}{{Eyer}, L.}, \bibinfo{year}{2023}.
\newblock \bibinfo{title}{{Gaia Data Release 3. The first Gaia catalogue of
  eclipsing-binary candidates}}.
\newblock \bibinfo{journal}{aap} \bibinfo{volume}{674}, \bibinfo{pages}{A16}.
\newblock \DOIprefix\doi{10.1051/0004-6361/202245330},
  \href{http://arxiv.org/abs/2211.00929}{\tt arXiv:2211.00929}.
%Type = Article
\bibitem[{{Mowlavi} et~al.(2017){Mowlavi}, {Lecoeur-Ta{\"\i}bi}, {Holl},
  {Rimoldini}, {Barblan}, {Pr{\v{s}}a}, {Kochoska}, {S{\"u}veges}, {Eyer},
  {Nienartowicz}, {Jevardat}, {Charnas}, {Guy} and {Audard}}]{Mowlavi2017}
\bibinfo{author}{{Mowlavi}, N.}, \bibinfo{author}{{Lecoeur-Ta{\"\i}bi}, I.},
  \bibinfo{author}{{Holl}, B.}, \bibinfo{author}{{Rimoldini}, L.},
  \bibinfo{author}{{Barblan}, F.}, \bibinfo{author}{{Pr{\v{s}}a}, A.},
  \bibinfo{author}{{Kochoska}, A.}, \bibinfo{author}{{S{\"u}veges}, M.},
  \bibinfo{author}{{Eyer}, L.}, \bibinfo{author}{{Nienartowicz}, K.},
  \bibinfo{author}{{Jevardat}, G.}, \bibinfo{author}{{Charnas}, J.},
  \bibinfo{author}{{Guy}, L.}, \bibinfo{author}{{Audard}, M.},
  \bibinfo{year}{2017}.
\newblock \bibinfo{title}{{Gaia eclipsing binary and multiple systems.
  Two-Gaussian models applied to OGLE-III eclipsing binary light curves in the
  Large Magellanic Cloud}}.
\newblock \bibinfo{journal}{aap} \bibinfo{volume}{606}, \bibinfo{pages}{A92}.
\newblock \DOIprefix\doi{10.1051/0004-6361/201730613},
  \href{http://arxiv.org/abs/1703.10597}{\tt arXiv:1703.10597}.
%Type = Article
\bibitem[{{Paegert} et~al.(2022){Paegert}, {Stassun}, {Collins}, {Pepper},
  {Torres}, {Jenkins}, {Twicken} and {Latham}}]{Paegert2022}
\bibinfo{author}{{Paegert}, M.}, \bibinfo{author}{{Stassun}, K.G.},
  \bibinfo{author}{{Collins}, K.A.}, \bibinfo{author}{{Pepper}, J.},
  \bibinfo{author}{{Torres}, G.}, \bibinfo{author}{{Jenkins}, J.},
  \bibinfo{author}{{Twicken}, J.D.}, \bibinfo{author}{{Latham}, D.W.},
  \bibinfo{year}{2022}.
\newblock \bibinfo{title}{{VizieR Online Data Catalog: TESS Input Catalog
  version 8.2 (TIC v8.2) (Paegert+, 2021)}}.
\newblock \bibinfo{journal}{VizieR Online Data Catalog} ,
  \bibinfo{pages}{IV/39}.
%Type = Article
\bibitem[{{Pols} et~al.(1998){Pols}, {Schr{\"o}der}, {Hurley}, {Tout} and
  {Eggleton}}]{Pols1998}
\bibinfo{author}{{Pols}, O.R.}, \bibinfo{author}{{Schr{\"o}der}, K.P.},
  \bibinfo{author}{{Hurley}, J.R.}, \bibinfo{author}{{Tout}, C.A.},
  \bibinfo{author}{{Eggleton}, P.P.}, \bibinfo{year}{1998}.
\newblock \bibinfo{title}{{Stellar evolution models for Z = 0.0001 to 0.03}}.
\newblock \bibinfo{journal}{mnras} \bibinfo{volume}{298},
  \bibinfo{pages}{525--536}.
\newblock \DOIprefix\doi{10.1046/j.1365-8711.1998.01658.x}.
%Type = Book
\bibitem[{{Pr{\v{s}}a}(2018)}]{Prsa2018}
\bibinfo{author}{{Pr{\v{s}}a}, A.}, \bibinfo{year}{2018}.
\newblock \bibinfo{title}{{Modeling and Analysis of Eclipsing Binary Stars; The
  theory and design principles of PHOEBE}}.
\newblock \DOIprefix\doi{10.1088/978-0-7503-1287-5}.
%Type = Article
\bibitem[{{Slawson} et~al.(2011){Slawson}, {Pr{\v{s}}a}, {Welsh}, {Orosz},
  {Rucker}, {Batalha}, {Doyle}, {Engle}, {Conroy}, {Coughlin}, {Gregg},
  {Fetherolf}, {Short}, {Windmiller}, {Fabrycky}, {Howell}, {Jenkins}, {Uddin},
  {Mullally}, {Seader}, {Thompson}, {Sanderfer}, {Borucki} and
  {Koch}}]{Slawson2011}
\bibinfo{author}{{Slawson}, R.W.}, \bibinfo{author}{{Pr{\v{s}}a}, A.},
  \bibinfo{author}{{Welsh}, W.F.}, \bibinfo{author}{{Orosz}, J.A.},
  \bibinfo{author}{{Rucker}, M.}, \bibinfo{author}{{Batalha}, N.},
  \bibinfo{author}{{Doyle}, L.R.}, \bibinfo{author}{{Engle}, S.G.},
  \bibinfo{author}{{Conroy}, K.}, \bibinfo{author}{{Coughlin}, J.},
  \bibinfo{author}{{Gregg}, T.A.}, \bibinfo{author}{{Fetherolf}, T.},
  \bibinfo{author}{{Short}, D.R.}, \bibinfo{author}{{Windmiller}, G.},
  \bibinfo{author}{{Fabrycky}, D.C.}, \bibinfo{author}{{Howell}, S.B.},
  \bibinfo{author}{{Jenkins}, J.M.}, \bibinfo{author}{{Uddin}, K.},
  \bibinfo{author}{{Mullally}, F.}, \bibinfo{author}{{Seader}, S.E.},
  \bibinfo{author}{{Thompson}, S.E.}, \bibinfo{author}{{Sanderfer}, D.T.},
  \bibinfo{author}{{Borucki}, W.}, \bibinfo{author}{{Koch}, D.},
  \bibinfo{year}{2011}.
\newblock \bibinfo{title}{{Kepler Eclipsing Binary Stars. II. 2165 Eclipsing
  Binaries in the Second Data Release}}.
\newblock \bibinfo{journal}{aj} \bibinfo{volume}{142}, \bibinfo{pages}{160}.
\newblock \DOIprefix\doi{10.1088/0004-6256/142/5/160},
  \href{http://arxiv.org/abs/1103.1659}{\tt arXiv:1103.1659}.
%Type = Article
\bibitem[{{van Roestel} et~al.(2018){van Roestel}, {Kupfer}, {Ruiz-Carmona},
  {Groot}, {Prince}, {Burdge}, {Laher}, {Shupe} and {Bellm}}]{vanRoestel2018}
\bibinfo{author}{{van Roestel}, J.}, \bibinfo{author}{{Kupfer}, T.},
  \bibinfo{author}{{Ruiz-Carmona}, R.}, \bibinfo{author}{{Groot}, P.J.},
  \bibinfo{author}{{Prince}, T.A.}, \bibinfo{author}{{Burdge}, K.},
  \bibinfo{author}{{Laher}, R.}, \bibinfo{author}{{Shupe}, D.L.},
  \bibinfo{author}{{Bellm}, E.}, \bibinfo{year}{2018}.
\newblock \bibinfo{title}{{Discovery of 36 eclipsing EL CVn binaries found by
  the Palomar Transient Factory}}.
\newblock \bibinfo{journal}{mnras} \bibinfo{volume}{475},
  \bibinfo{pages}{2560--2590}.
\newblock \DOIprefix\doi{10.1093/mnras/stx3291},
  \href{http://arxiv.org/abs/1712.06507}{\tt arXiv:1712.06507}.
%Type = Article
\bibitem[{{Wang} and {Han}(2010)}]{Wang2010}
\bibinfo{author}{{Wang}, B.}, \bibinfo{author}{{Han}, Z.},
  \bibinfo{year}{2010}.
\newblock \bibinfo{title}{{Companion stars of Type Ia supernovae and single
  low-mass white dwarfs}}.
\newblock \bibinfo{journal}{mnras} \bibinfo{volume}{404},
  \bibinfo{pages}{L84--L88}.
\newblock \DOIprefix\doi{10.1111/j.1745-3933.2010.00840.x},
  \href{http://arxiv.org/abs/1002.4742}{\tt arXiv:1002.4742}.
%Type = Article
\bibitem[{{Wang} et~al.(2022){Wang}, {N{\'e}meth}, {Luo}, {Chen}, {Jiang} and
  {Cao}}]{Wang2022}
\bibinfo{author}{{Wang}, K.}, \bibinfo{author}{{N{\'e}meth}, P.},
  \bibinfo{author}{{Luo}, Y.}, \bibinfo{author}{{Chen}, X.},
  \bibinfo{author}{{Jiang}, Q.}, \bibinfo{author}{{Cao}, X.},
  \bibinfo{year}{2022}.
\newblock \bibinfo{title}{{Extremely Low-mass White Dwarf Stars Observed in
  Gaia DR2 and LAMOST DR8}}.
\newblock \bibinfo{journal}{apj} \bibinfo{volume}{936}, \bibinfo{pages}{5}.
\newblock \DOIprefix\doi{10.3847/1538-4357/ac847c},
  \href{http://arxiv.org/abs/2207.13401}{\tt arXiv:2207.13401}.
%Type = Article
\bibitem[{{Wang} et~al.(2023){Wang}, {Ren}, {Andersen}, {Grundahl}, {Chen} and
  {Pall{\'e}}}]{Wang2023AJ}
\bibinfo{author}{{Wang}, K.}, \bibinfo{author}{{Ren}, A.},
  \bibinfo{author}{{Andersen}, M.F.}, \bibinfo{author}{{Grundahl}, F.},
  \bibinfo{author}{{Chen}, T.}, \bibinfo{author}{{Pall{\'e}}, P.L.},
  \bibinfo{year}{2023}.
\newblock \bibinfo{title}{{FX UMa: A New Heartbeat Binary System with Linear
  and Nonlinear Tidal Oscillations and {\ensuremath{\delta}} Sct Pulsations}}.
\newblock \bibinfo{journal}{aj} \bibinfo{volume}{166}, \bibinfo{pages}{42}.
\newblock \DOIprefix\doi{10.3847/1538-3881/acdac9},
  \href{http://arxiv.org/abs/2212.04639}{\tt arXiv:2212.04639}.
%Type = Article
\bibitem[{{Wang} et~al.(2020a){Wang}, {Zhang} and {Dai}}]{Wang2020}
\bibinfo{author}{{Wang}, K.}, \bibinfo{author}{{Zhang}, X.},
  \bibinfo{author}{{Dai}, M.}, \bibinfo{year}{2020}a.
\newblock \bibinfo{title}{{Discovery of Two Pulsating Extremely Low-mass
  Pre-white Dwarf Candidates in the TESS Eclipsing Binaries}}.
\newblock \bibinfo{journal}{apj} \bibinfo{volume}{888}, \bibinfo{pages}{49}.
\newblock \DOIprefix\doi{10.3847/1538-4357/ab584c}.
%Type = Article
\bibitem[{{Wang} et~al.(2020b){Wang}, {Gies}, {Lester}, {Guo}, {Matson},
  {Peters}, {Dhillon}, {Butterley}, {Littlefair}, {Wilson} and
  {Maxted}}]{Wang2020AJ}
\bibinfo{author}{{Wang}, L.}, \bibinfo{author}{{Gies}, D.R.},
  \bibinfo{author}{{Lester}, K.V.}, \bibinfo{author}{{Guo}, Z.},
  \bibinfo{author}{{Matson}, R.A.}, \bibinfo{author}{{Peters}, G.J.},
  \bibinfo{author}{{Dhillon}, V.S.}, \bibinfo{author}{{Butterley}, T.},
  \bibinfo{author}{{Littlefair}, S.P.}, \bibinfo{author}{{Wilson}, R.W.},
  \bibinfo{author}{{Maxted}, P.F.L.}, \bibinfo{year}{2020}b.
\newblock \bibinfo{title}{{The Pre-He White Dwarf in the Post-mass Transfer
  Binary EL CVn}}.
\newblock \bibinfo{journal}{aj} \bibinfo{volume}{159}, \bibinfo{pages}{4}.
\newblock \DOIprefix\doi{10.3847/1538-3881/ab52fa}.
%Type = Article
\bibitem[{{Zhang} et~al.(2017){Zhang}, {Fu}, {Liu}, {Luo} and
  {Ren}}]{Zhang2017}
\bibinfo{author}{{Zhang}, X.B.}, \bibinfo{author}{{Fu}, J.N.},
  \bibinfo{author}{{Liu}, N.}, \bibinfo{author}{{Luo}, C.Q.},
  \bibinfo{author}{{Ren}, A.B.}, \bibinfo{year}{2017}.
\newblock \bibinfo{title}{{Low-mass Pre-He White Dwarf Stars in Kepler
  Eclipsing Binaries with Multi-periodic Pulsations}}.
\newblock \bibinfo{journal}{apj} \bibinfo{volume}{850}, \bibinfo{pages}{125}.
\newblock \DOIprefix\doi{10.3847/1538-4357/aa9577}.

\end{thebibliography}

% \appendix
% \section{APPENDIX: ADDITIONAL TABLES AND FIGURES}

\begin{figure*}
   \vspace{-0.5cm}
   \centering
   \includegraphics[width=14cm, angle=0]{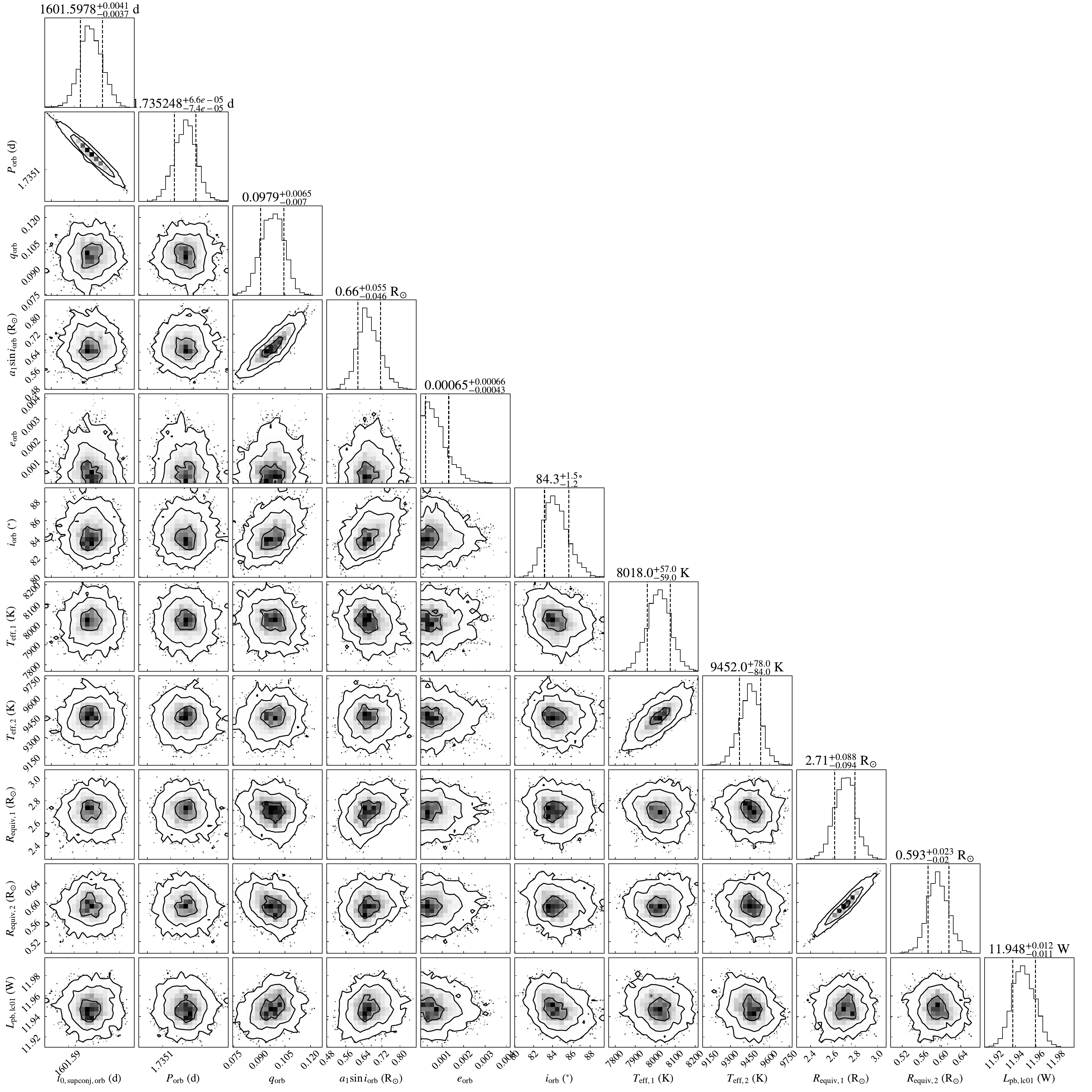}
   \caption{The posterior distributions of the binary parameters optimized in the PHOEBE fits to the TIC 100011519.}
   \label{fig5}
\end{figure*}

\begin{figure*}
\centering
\includegraphics[width=14cm, angle=0]{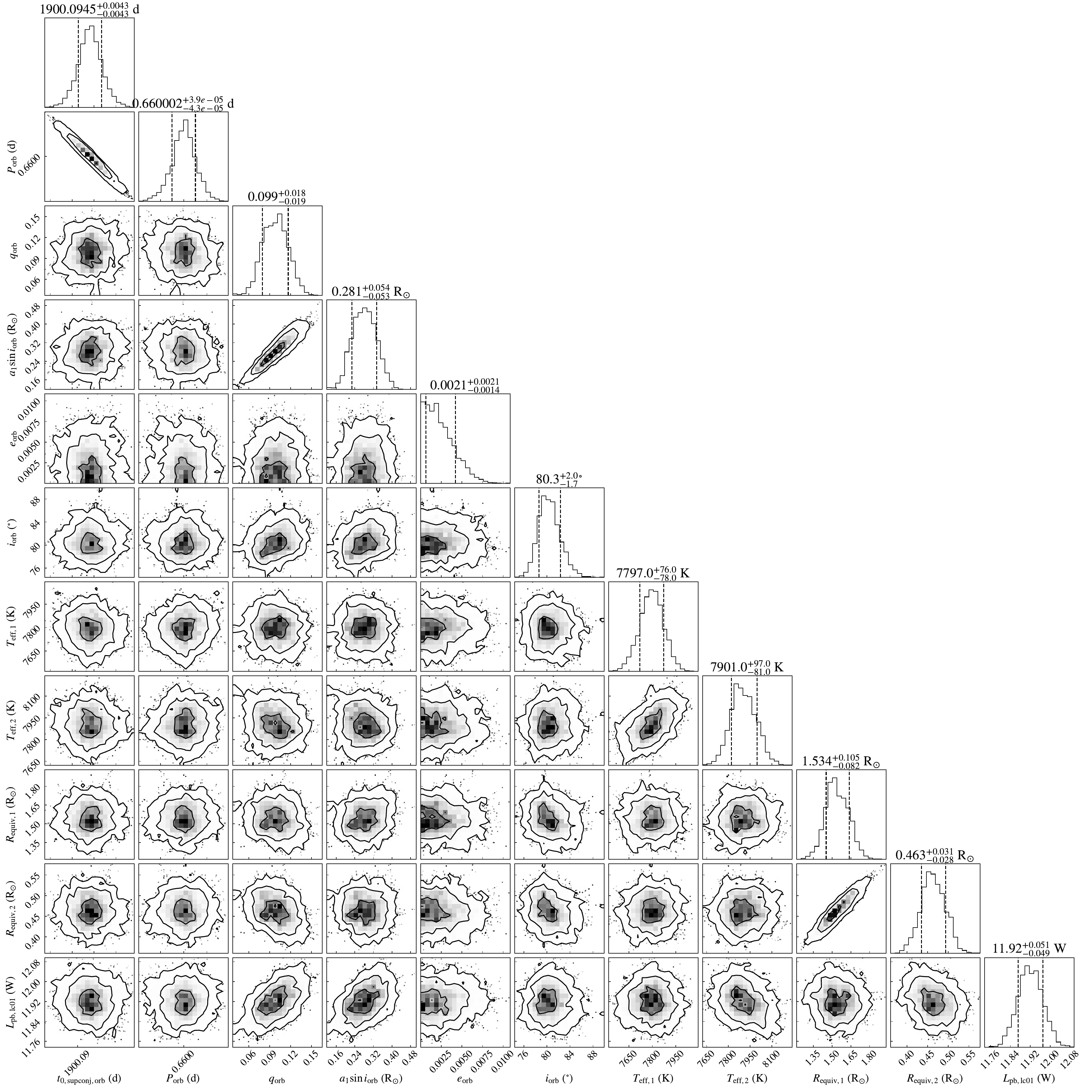}
\caption{The posterior distributions of the binary parameters optimized in the PHOEBE fits to the TIC 219485855.}
\label{fig6}
\end{figure*}

\begin{figure*}
\centering
\includegraphics[width=14cm, angle=0]{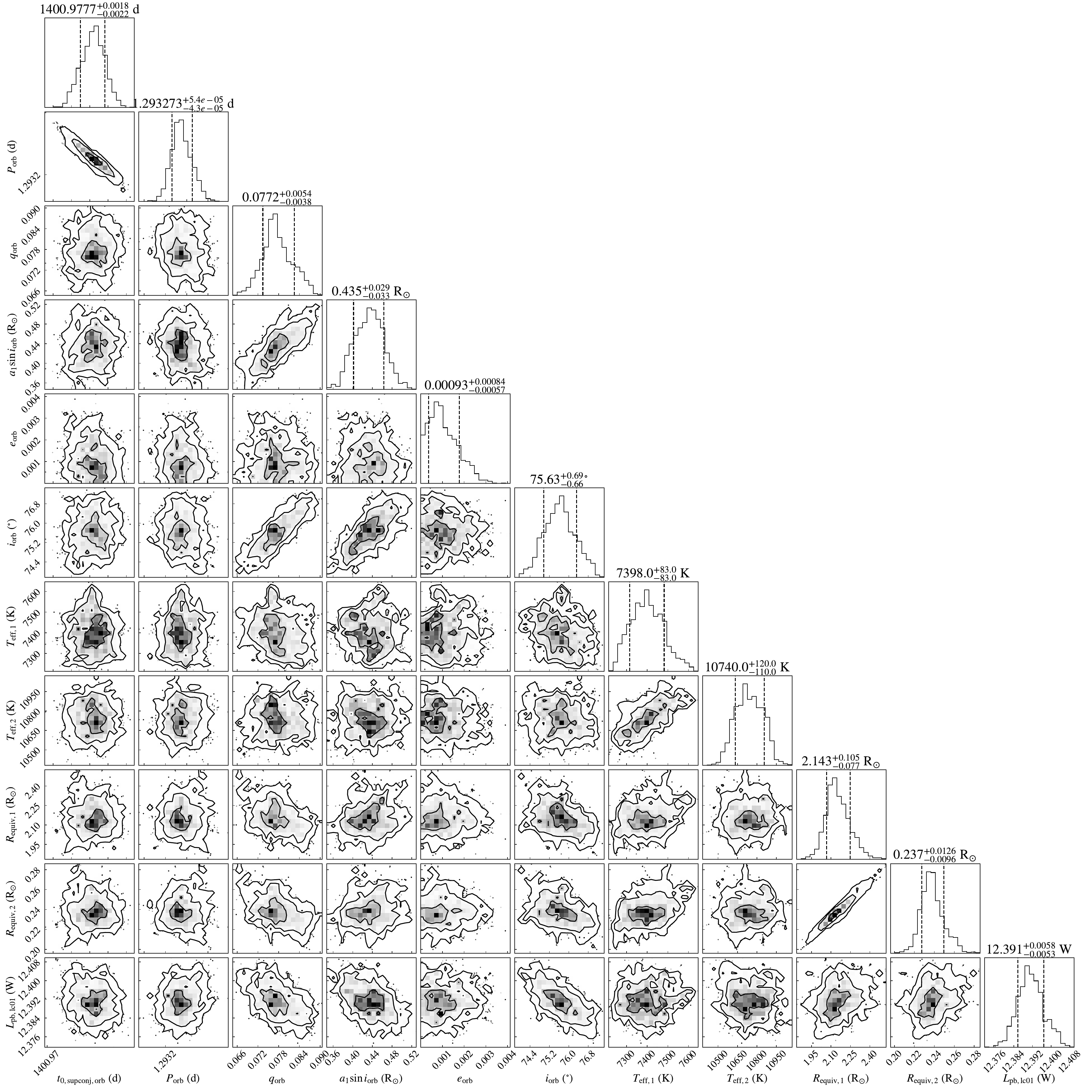}
\caption{The posterior distributions of the binary parameters optimized in the PHOEBE fits to the TIC 399725538.}
\label{fig7}
\end{figure*}

\begin{figure*}
\centering
\includegraphics[width=14cm, angle=0]{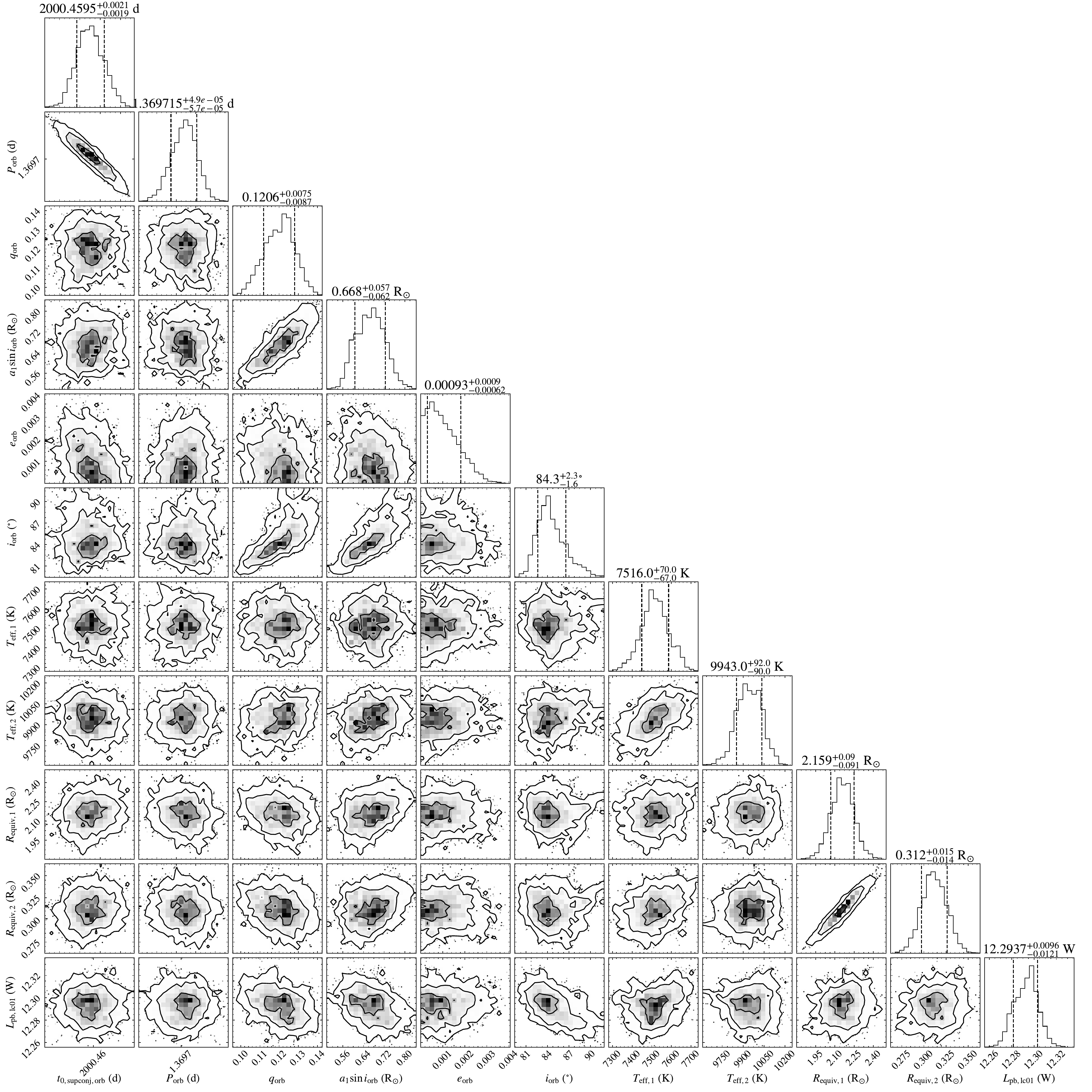}
\caption{The posterior distributions of the binary parameters optimized in the PHOEBE fits to the TIC 464641792.}
\label{fig8}
\end{figure*}

\end{document}